# Neuroscience and Literacy: An Integrative View

George Ellis (Mathematics Department, University of Cape Town)[1]

Carole Bloch (Education Department, University of the Western Cape, and PRAESA)[2]

**Abstract** *Significant challenges exist globally regarding literacy teaching and learning. To address these challenges, key features of how the brain works should be taken into account. First, perception is an active process based in detection of errors in hierarchical predictions of sensory data and action outcomes. Reading is a particular case of this non-linear predictive process. Second, emotions play a key role in underlying cognitive functioning, including oral and written language. Negative emotions undermine motivation to learn. Third, there is not the fundamental difference between listening/speaking and reading/writing often alleged on the basis of evolutionary arguments. Both are socio-cultural practices that are driven through the communication imperative of the social brain. Fourth, both listening and reading are contextually occurring pyscho-social practices of understanding, shaped by current knowledge and cultural contexts and practices. Fifth, the natural operation of the brain is not rule-based, as is supposed in the standard view of linguistics: it is prediction, based on statistical pattern recognition. This all calls into question narrow interpretations of the widely quoted "Simple View of Reading", which argues that explicit decoding is the necessary route to comprehension. One of the two neural routes to reading does not involve such explicit decoding processes, and can be activated from the earliest years. An integrated view of brain function reflecting the non-linear contextual nature of the reading process implies that an ongoing focus on personal meaning and understanding from the very beginning provides positive conditions for learning all aspects of reading and writing.*

## 1.      Introduction: The context for this paper

The well-known global debates and divisions in relation to the way children become literate have increasingly been influenced by neuroscience evidence on reading (Seidenberg *et al* 2020). Research undertaken by academics and researchers in the field, concentrated in the powerful centres of the Global North (especially the USA and the UK), have far reaching impact on those making policy and developing national documents (sometimes legally framed) at government level in relation to literacy teaching in diverse settings (Hoffman 2012).  This in turn influences pedagogy, curriculum approaches and teaching methods. The longstanding proposed 'Science of Reading'(Shanahan 2020, Seidenberg *et al* 2020) has increasingly been used by some involved in these debates to claim  that neuroscience studies support the *Simple View of Reading* (SVR),  a model of  reading which views the reading process in unidirectional linear terms. Literacy teaching debates and practices are at present dominated globally by approaches arising from the SVR  (see Castles *et al* 2018, Clark 2020 and a special issue of the *Reading Research Quarterly*: Goodwin and Jiménez 2020).  Because of the claimed supporting neuroscientific evidence, there is now a widely perceived gravitas and authority adhering to this particular model, which insists on building skills as a prior step to comprehension. It is used to claim that the 'reading wars' which pitted phonics against whole language should now be over (Castles *et al* 2018). Many literacy specialists and teachers find an either-or position misleading and unhelpful; some prefer a 'Balanced Approach' as a middle-ground to ensure children get 'the best of both worlds' in teaching programs (Willson and Falcon 2018). In any event there is no consensus over common narrow interpretations of the SVR; several recent papers focus on the complex and multifaceted interplay between decoding and listening comprehension (Cervetti *et al* 2020, Compton-Lilly *et al* 2020, Bua Lit Collective 2018) or even propose a *Complete View of Reading* (CVRi) (Francis *et al* 2018) or similar (Snow 2018).

---

[1] email: george.ellis@uct.ac.za
[2] email: cbloch@uwc.ac.za

Still, aspects of reductive neuroscience are used to justify the kind of singular teaching focus on skills, which is currently commonly viewed as necessary for all children as they begin their formal schooling. This is understood to be the case irrespective of vastly diverse socio-economic and cultural contexts and individual experiences, with immense significance for the serious global literacy teaching challenges. South Africa is a case in point. The longstanding systemic problems with teaching early literacy effectively in this particular context (Taylor 1989, Bloch 1999, 2000, Alexander and Bloch 2010, NEEDU 2013), characterised by historical language inequities and extremes of economic and social inequality, illustrates how urgent this matter is. Several government led initiatives have been attempted to improve matters since apartheid ended (e.g. DoE 2008, DoE 2011, Van der Berg *et al* 2016), but children have continued to perform badly on all assessments; and particularly on PIRLS, which is focused on comprehension. The 2016 results claimed that 78% of South African children could not read for meaning by the end of grade 4 in African languages or English (Howie et al. 2017). This intensified efforts and discussions to both understand why this situation exists, and to provide viable solutions to get children 'reading for meaning by age 10'[3] (Reeves 2017, Hickman, 2018, Bua-lit Language and Literacy Collective 2018, Fleisch and Dixon 2019). But the view of reading which dominates research, policy and curriculum related interventions reflects the Simple View of Reading (Spaull *et al* 2020) and early literacy teaching in classrooms is still characterised by the foregrounding of decontextualised exercises.

~~Recent initiatives include~~ an acknowledgement that learning to read and write should begin in languages children understand; this has led to attention being given to establishing reading benchmarks in African languages (Spaull *et al* 2020, Jukes *et al* 2020); also increased government level action on the recognition that the years before formal school are crucial ones for laying firm learning foundations (DoE 2015, Harrison 2020). This is in response to the broad global consensus that during their preschool years, children learn best in informal, meaning and play based ways, including their oral language foundations and first steps to literacy. Yet there is an increasing push down pressure to consider earlier skills teaching from formal schooling, compromising time for play in the preschool years (Campbell 2020). And, while the curriculum for the first 4 years of primary schooling also orients towards meaningful teaching and learning (DoE 2011) a conceptual schism exists between curriculum statements and their teaching implications on the one hand and on the other, popular understandings and practices regarding what to prioritise for initial literacy teaching in school. Grade R, which is simultaneously the last preschool year and /or the first primary school year, is caught at the centre of this conceptual and practical conundrum.

---

[3] These terms became widely used in South Africa since the President of South Africa called for all children to 'read for meaning' by aged 10 in his 2019 State of the Nation Address, after being alerted to the severity of the challenges following these results.

In the interests of working towards equity and justice for all children, our concern in this paper is to problematise the validity of the reductive neuroscience view of how reading works. We do this by offering alternative evidence from the growing body of integrative neuroscience which perceives reading as a non-linear holistic process strongly influenced by affect and involving a foundational search for meaning through prediction, developing through tentative exploration as skills are built. In particular, we dispute the widely claimed view that oral language is natural but written language is not (Shaywitz 2003:49-50, Wolf 2018). Rather we argue that both are cultural inventions that originated at different times through similar evolutionary processes in the long distant past, in order to meet social needs (Harari 2011).

**A note on terminology** In the educational body of literature on teaching literacy, the term 'reading' has been used far more than 'writing'. This reflects how these aspects of literacy tend to be viewed and taught separately. More recently, the term 'literacy' is being used as a conscious umbrella term to bring more integrative socio-cultural understandings to bear (Frankel *et al* 2016). In this paper, when referring to published work we tend to use the term 'reading' as the authors often do, while in our own writing we use 'writing and reading' and 'literacy' synonymously, unless we are specifically referring to one of them.

### 1.1 Contrasting early literacy perspectives

The different views about literacy and how it is learnt arose from historical disagreements about the nature of knowledge and how language learning happens (Altwerger *et al* 2007:4). Two contrasting pedagogical perspectives co-exist, and have been argued about for hundreds of years (Huey 1908, Chall 1967, Pearson 2004, Kim 2008, Castles *et al* 2018, Miller 2020). The central issue has come to be how and when comprehension comes about. These two perspectives can be related to two models of literacy (Street 1984). One, which Street (2006) calls the 'autonomous model', views literacy as constituting separate sets of skills, to be taught independent of context. The other views literacy as being based in the social practices of communities. In this 'ideological model', there are different forms and uses for literacy in different socio-cultural and linguistic settings, and these form the basis for teaching. Understandings about young children's literacy learning and teaching can be viewed as falling under one or other of these umbrellas.

Early literacy teaching approaches which correspond to Street's broad autonomous model of literacy are underpinned by '*skills based*'[4] models (summarised in **Figure 1**). Now supported by reductive neuroscience studies based in a linear view of cognition and action, the automatic decoding of phonics skills (recognising and sounding out letter- sound relationships) followed by fluency are widely seen as essential prior steps to comprehension, because of how the brain is believed to function. This applies to each language a child is being taught to read. The '*meaning based*'[5] model (summarised in **Figure 2**), which corresponds to Street's ideological model, underpins approaches which see initial and ongoing meaning construction taking place, with alphabetic knowledge and phonics skills being taught in the context of authentic literacy related experiences. This implies using relevant languages[6] and a focus on motivation, personal agency, and meaning, with predictive understandings connecting content to children's socio-cultural understandings and practices.

---

[4] We use the terms *skills based* to refer to 'part to whole' views of early literacy teaching which, may differ in detail, but all start from the 'bottom up' with phonics and other technical skills, also referred to as 'phonics based', or Structured Literacy.

It is important to note that in a meaning-based model[7] there is no 'profound mistake' being made through omitting phonics, as alleged by Seidenberg *et al* (2020); because the view is of a complex meaning-based process with grapho-phonic cues working in concert with semantic and syntactic cues, as indicated in Figure 2. Phonics is taught as it arises in the texts being read and also as it is required to write.

**Informal learning before school**: A significant body of international interdisciplinary early literacy research evidence has been conducted into the years before formal schooling.[8] Those with meaning-based perspectives tend to conceive of learning related to written language as forming part of the informally structured foundations of learning processes, consistent with early childhood wisdoms and traditions which value holistic learning and play (Bruce 2015). The view is that these foundations ought to be deepened and expanded as school begins (Bua lit 2018). By contrast, from a skills based perspectives, such learning is usually conceptualised as preparatory: pre-reading and pre- writing activities, the basic building blocks which are needed to be taught to young children so that they are ready for the formal teaching of reading and writing in school. The strong implication is that proper literacy learning begins here.

**Formal learning in school:** At the start of formal schooling, many teaching programmes follow the narrow skills based interpretation of the SVR (Gough & Tunmer 1986, Compton-Lilly *et al* 2020). In doing so, they may neglect to emphasise and enable crucial meaning-based elements and experiences children require in the vital early stages of becoming literate, thereby restricting opportunities for appropriate quality learning. We argue that from the early years onwards, major features of how the predictive brain works, which are currently ignored in the early literacy teaching literature, need to be taken into account in order to ensure appropriate conditions of learning (Cambourne 1995, 2000, 2020) for all young children whenever they encounter written language. We do not agree that these conditions are different for children from low SES, poorly served communities (Abadzi 2006, 2008).

---

[5] We use the term *meaning based* to refer to views of early literacy teaching which may differ in detail, but prioritise context, socio-cultural practices and meaning making. They are sometimes called 'top down', emergent literacy, whole language or 'social practices'.
[6] Although we do not deal directly with multilingualism, multiliteracies, and learning in this paper, we flag this as involving significant pedagogical issues which are impacted on directly by the views of neuroscience which underpin programs for language and literacy teaching for all children.
[7] We avoid the term "whole language" as this is a loaded term, with various interpretations and misinterpretations of its meaning.
[8] To date, most of this early literacy research has been done in high Socio-Economic Status (SES) countries of the Global North. Despite significant recent scientific attention on the importance of the 'first 1000 days', the early years of childhood are still poorly provided for, and are very low in actual status and value in terms of support for quality care and educational provision, except for the children of the elite. Slowly it is being instituted in the Global South; research attention follows in its trail.

**1.2 Understanding based in reductive neuroscience views**

Many studies present the brain regions involved in oral language (Friedericki 2017) and written language (Shaywitz 2003, Dehaene 2010, Wandell *et al* 2012, Kearns *et al* 2019). Some psychologically or cognitive based texts give a brief presentation of the neuroscience, e.g., Deacon (1998), Wolf (2008), Schnelle (2010), Seidenberg (2017), Hruby and Goswami 2019). Others model cognitive processes without linking to neuroscience proper, e.g., Tomasello (2003), Willlingham (2017). Many link language to evolution, e.g., Tomasello (2000), Donald (2001), Greenspan and Shanker (2004). We note that much of the literature which focuses specifically on neuroscience and reading has grown out of studies related to dyslexia, e.g., Shaywitz (2003) and Wolf (2008). An important question is thus to what extent studies of dyslexia throw light on normal reading processes?[9]. These studies have by definition a deficit view of the reading process built in, which means their recommendations will necessarily be affected by that view[10].

In many of these writings, the link to neuroscience is limited to diagrams of active domains and pathways in the brain when phonemes, words, or non-words are read, or more accurately, decoded, particularly referring to the Visual Word Form Area (VWFA).[11] These are supported by functional neuroimaging studies. While this gives useful information about neural pathways associated with reading, one should be aware that they are rarely accurate representations of the full functional brain networks operating when a person reads or attempts to read meaningful texts, and hence they only give a very partial picture of what goes on in the brain when such purposeful reading takes place.[12] Furthermore many assume that perception operates in a linear manner from sensory data input to analysis of that data in the cortex, resulting in the SVR. For this reason, we term this reductionist neuroscience.

To focus the discussion, we refer mainly to four bodies of work which encapsulate the reductionist neuroscience view: Shaywitz (2003) because her neuroscience research into dyslexic children's brains informs 'normal' reading. as do her views of natural and unnatural language; Dehaene (2010), as it is in many ways the ground work on reading and the brain that many others refer back to**;** Abadzi (2006, 2008, 2017), who has been immensely influential for development aid literacy programmes via her work at the

---

[9] We use the term 'normal' here to include diverse SES, cultural and linguistic practices and contexts.

[10] It is pertinent to consider how this might affect both the confidence of young beginning readers who live in poorly served communities and expectations of them as readers. Many are taught by teachers who have been trained to perceive them as already lacking in school readiness skills; a deficit model of reading is added to this.

[11] The role of the VWFA as unique to reading has been called into question recently, inter alia by Vogel (2012, 2014), Moore (2014), Martin (2019), and Vidal *et al* (2021). The VWFA is not present on functional MRI scans before learning to read, but appears and enlarges as reading skill is gained. Also noted, is that from the outset the VWFA is strongly connected to the dorsal tempero-parietal areas which are activated during speaking and listening, and are present as early as 2 months of age. The fMRI scans show that, after the initial visual reception, reading results in nearly instantaneous and simultaneous involvement of widespread ventral, dorsal and frontal areas involved in the sound, shape and meaning of words in skilled readers. We thank Roland Eastman for these comments.

[12] A study which does this is Fedorenko *et al* (2016), showing how different brain areas are indeed involved.

World Bank and The Global Partnership for Education in the Global South;[13] and Castles *et al* (2018), as this paper summarises the SVR and is an up to date review of the "reading wars" between the two positions on literacy and how it should be taught;.

Castles *et al* (2018) summarise the *Simple View of Reading* (SVR) (Gough and Tunmer, 1986) thus:
> "The Simple View of Reading posits that reading comprehension R is the product of two sets of skills, `decoding' D and `linguistic comprehension' C :
> $$R = D \times C. \qquad (1)$$
> The logical case for the Simple View is clear and compelling: Decoding and linguistic comprehension are both necessary, and neither is sufficient alone. A child who can decode print but cannot comprehend is not reading; likewise, regardless of the level of linguistic comprehension, reading cannot happen without decoding. ... Early in development, reading comprehension is highly constrained by limitations in decoding. As children get older, the correlation between linguistic and reading comprehension strengthens, reflecting the fact that once a level of decoding mastery is achieved, reading comprehension is constrained by how well an individual understands spoken language."

They then use this view that comprehension is initially constrained by limitations in decoding to motivate the imperative of a skills based model (although they do also emphasize the importance of broader reading experiences). But firstly, decoding as such is only necessary for the Dorsal Pathway, one of the two neural reading pathways that they describe (see Section 5.3 below): it does not explicitly occur in the Ventral Pathway because graphophonic as well as other structural language features are always sampled to the extent that they are needed to predict the meaning of the text. Miscue analysis (Flurkey *et al* 2008) demonstrates that successful reading involves preserving meaning of the text, though not necessarily word accuracy. So letter by letter decoding is not in fact necessary in order to read: words can be grasped as a whole in a *gestalt* way (Section 2.1). Secondly, comprehension early in development is more likely to be constrained when reading is taught with the strong primary emphasis on decoding skills they recommend. This restricts the child's attempts to understand the text directly by drawing on other clues, because of the way teaching focuses the child's attention towards accuracy and fast decoding. For example Spear-Swerling (2019) argues against encouraging students to attend to multiple-cueing systems[14] when reading, which is what a mature reader will do.

Dehaene makes this explicit when he says
> "*The child's brain, at this stage, is attempting to match the general shape of the words directly onto meaning, without paying attention to individual letters and their pronunciation – a sham form of reading*" (Dehaene 2010:200).

---
[13] Klaas and Trudell (2010), Piper *et al* (2016), and including South Africa, see for example Spaull & Pretorius (2016:9).
[14] We refer to multiple cueing systems below in Section **5.2.**

His is acknowledging that the use of the ventral pathway is possible and indeed young children can do so, but his reductionist perspective leads him to recommend preventing this from happening. He defines reading inadequately. He wants the parts to work rather than the integral process, and characterizes as 'sham' reading that which is both the intention of proficient readers and a profound reading path for young learners. He shuns precisely what children need to do to avoid a possible memory overload, which is a reason given for the need to concentrate children's attention on developing swift and automatic decoding (see Section 5.4). Dehaene also dissuades teachers from encouraging children from making attempts at conventional reading:

> *"Children need to understand that only the analysis of letters one by one will allow them to discover a word's identity"* (Dehaene 2010:229).

This contradicts the predictive understanding of perception we highlight below, see for example Friston *et al* (2017a), and ignores the other cueing systems which proficient readers use and which should be encouraged and supported in learners. The serious problem is that Dehaene's authoritative advice to educators (Dehaene 2010:230), where everything is planned to the last grapheme, is a recipe for rigidity that makes no allowance for prior knowledge and development and social and cultural experiences, or the role of motivation and the drive towards understanding. He makes statements against including illustrations in books (Dehaene 2010:229) or posters on the wall. This bleak view of early literacy teaching completely ignores the powerful symbolic life and imagination of young children, their impressive linguistic and intellectual capabilities, and the affective dimension of the mind that we emphasize below.

Abadzi (2017:8) offers a view of comprehension where she claims that, in contrast to its usual prominent position in high SES educational contexts, "comprehension" need not be the aim of the learning process for poor children (Abadzi: 2017:8):

> *"Should instruction focus on reading comprehension early on? Middle-class children often process quickly and have rich vocabulary; so, in high-income countries, literal comprehension may be too simplistic. Instead, ''comprehension'' is often used to signal inferences or predictions. These require more knowledge than offered in a text. Poorer students have more limited vocabulary and expression, and they may lack the academic language to deal with classroom conversations."*

This view suggests reading need not imply comprehension, and that is precisely the problem that can occur when the focus is on teaching skills out of context. Does she believe that it's 'natural' for middle-class children to develop rich vocabularies? This highly problematic position begs the question of how to address deep inequalities in transformative ways to promotes intellectual and affective justice and equity. She continues:

> *"To teach the poor efficiently, we must make learning easiest on their brains. The research suggests that, when time is scarce, reading components could be taught sequentially. The sequence could roughly follow that of the reading stimuli as they go through the brain. Teachers must focus instruction and practice on the early visual processes and speed those up in order to facilitate*

> *complex cognition. Middle-class reading instruction, such as the simultaneous teaching of the ''five pillars''[15] may slow down and complicate the acquisition of this quintessentially visual skill. The answer to the twenty-first-century reading crisis may lie in second-century practices, such as decoding, that apparently most human brains could perform"* (Abadzi 2017: 11).

Visual processes are not the bottleneck, because of the predictive nature of vision which involves the cortico-thalamic feedback circuits. Much information needed for interpretation is already present before the signal arrives. The extreme position taken by Abadzi implies that the brains of poor (African language speaking?) children are different from those of more affluent ones and are unable to deal with complexity. Apart from being insulting and patronising, it misleads teachers and learners down imaginative and intellectual *cul de sacs*. In Abadzi (2008) she makes a major issue out of the mind's short-term timeframe and the need to read fast. But there is no need to read fast, the need is to read and to learn to read with comprehension (Dowd and Bartlett 2019). The problem arises if one insists that "*Reading starts with tracking and interpreting individual letters in a morass of print*" (Abadzi 2008) and then teach in such a way as to enforce this as the priority. She later admits that "*Fluency is achieved when an instant word recognition pathway is activated*" (this is the ventral pathway). She claims "*This happens after much practice in pairing consistently sounds with groups of letters*".

However teaching approaches which emphasise the authentic language of storytelling expose children to precisely the rich vocabulary and expression which Abadzi claims they may not have. In settings which have been dominated by colonial and post-colonial education systems, reinstating story as a legitimate educational form is worthy in itself, as well as providing an obvious segue to written language; Motivation comes as adults and children connect with personal and cultural histories, at the same time as they create some of the texts to read.

Shaywitz emphasizes the view that oral language is natural, and written language is unnatural, for example
> "*Spoken language is instinctive, built into our genes and hardwired into our brains. Learning to read demands that we take advantage of what nature has provided: a biological model for language*" (Shaywitz and Shaywitz (2004).

She, and many others, have used this as one of the powerful motivators for skills-based reading models (e.g. Wolf 2018, Spaull and Pretorius 2019:5). We strongly critique this understanding in Section 4 below.
.

**1.3 Understanding based in integrative neuroscience views**

<u>Miłkowski *et al* (2018) claim that</u> cognitive neuroscience has undergone a silent revolution based in the integration of wide perspectives with the rest of the cognitive neurosciences. These substantial change in neuroscience perspectives on brain function develop from earlier views on how perception works, for example Gombrich (1961), Gregory (1978), and Purves (2010), leading to the hierarchical predictive processing view of action and perception espoused by Friston (2003, 2010, 2012), Clark (2013, 2016), Hohwy (2013), Seth (2013), Fabry (2017), and many others, giving a more integrative view of brain function.  In discussing this integrative neuroscience and its relevance for literacy learning and teaching, we point out the importance of five major features of how the brain works:

- First, perception is an active, contextually based predictive process, based in detection of errors in hierarchical predictions of sensory data and action outcomes. Reading and writing are particular cases of this process. Not all text need be read; words can be filled in due to context **(Figure 3)**.
- Second, emotions play a key role in underlying cognitive functioning. Innate affective systems underlie and shape all brain functioning, including communicating by speech and writing.
- Third, there is not the fundamental difference between listening/speaking and reading/ writing that is often alleged on the basis of evolutionary arguments. They are both social and cultural practices learnt through social processes.
- Forth, brain function is not fundamentally based in a rule-based way of responding to data. It is a neural network of huge dimensions, whose natural mode of operation is statistical pattern recognition and prediction, based in non-local storage of data. It is a Bayesian machine.
- Fifth, like listening, reading is a non-linear contextually shaped psycho-social process of conveying meaning in a specific context, shaped by current knowledge. One of the two neural routes to reading does not involve explicit decoding processes, and can be activated from the earliest years.

---
[15] These 5 pillars were identified by the National Reading Panel (2000) as phonemic awareness, phonics, reading fluency, vocabulary and reading comprehension.

This predictive nature of perception is enabled by cortico-thalamic circuitry (Allitto and Usrey 2003) allowing downward passing of predictions from the cortex to the thalamus, as depicted in **Figure 4**.  We contend that a 21st century perspective on literacy must include this evidence about not just uni-directional but bi-directional  neural messages passing in hierarchical systems.  This processing is affected by affective (emotional) messages passed diffusely from the limbic system to the neocortex via ascending systems (**Figure 5**). Crucially, there are two neural routes to reading  – the "indirect" (dorsal) and "direct" (ventral) pathways (**Figure 6**). We argue that the direct path is a powerful biologically natural way by which beginning readers can learn to read without having to explicitly decode. The resulting view of the reading process corresponds  with the meaning-based views proposed *inter alia* by Goodman (1967, 1982), Strauss *at al* (2009), Bever (2009, 2013, 2017), and Goodman *et al* (2016). .

In what follows, we discuss each of the five major features in detail. **Section 2** looks at the brain and how perception is an active process, **Section 3** at how cognitive function is crucially shaped by affect (emotions), **Section 4** looks at what is `natural': what are the innate brain systems?, **Section 5** at how the natural mode of operation of the brain is statistical pattern recognition and prediction, and **Section 6** considers the similar neural and psychological processes involved in meaning making and communicating by listening/speaking and reading/writing. **Section 7** comments briefly on possible educational implications.

## 2      Perception is an active, contextually based predictive process

The brain works in a complex, non-linear way. The neocortex is a predictive organ (Hawkins, 2005 Kveraga *et al* 2007) based in connectionist principles (**Section 5**). It is the seat of perception and pattern recognition, learning based in neural plasticity, and sensation/action based in prediction and choice (Purves *et al* 2008, Gray 2011). Downward causation takes place in a variety of ways: in relation to perception, attention, and motor control (Ellis 2016, 2018). Reading and listening are forms of perception; speaking and writing are forms of action modulated by perception.

**2.1 How perception works: Hierarchical predictive Processing**
The key point we wish to raise here is the predictive way all sensory systems work as discussed by Gregory (1978) and many others. The brain understands the world in a holistic way on the basis of the clues offered to it (Purves 2010, Kandel 2016). It has to do this in order to solve Helmholz's inverse problem, namely we are not provided by our senses with enough data to uniquely determine what the situation 'out there' is. We have to do the best we can with what sensory data is available, even though some needed data is missing. Consequently vision is an active process (Findlay and Gilchrist 2003).

Like vision (Frith 2007, Purves 2010), reading involves prediction in the light of previous experience and the confirmation or adjustments of such predictions in the light of new information (Smith 2004). This is nothing other than the process of hierarchical predictive processing[16] (Friston 2003, 2010, Clark 2013, Hohwy 2013, Seth 2013, 2014), which underlies how reading text with meaning actually takes place (Goodman 1967, Smith 2004, Flurkey *et al* 2008, Strauss *at al* 2009, Goodman *et al* 2016). This is indicated by eye-tracking and miscue studies[17] as well as our ability to read scrambled or partially constituted pieces of text.

All perception works in the same contextual way because they are all based in the same cognitive mechanism, applied in different domains. They all proceed by in advance predicting what ought to be perceived, and then adjusting the predictions on the basis of incoming data (Bever and Poeppel 2010, Yon 2019) so as to minimise surprise (Friston 2010). This is stated by Clark (2013) as follows:

> *"Brains, it has recently been argued, are essentially prediction machines. They are bundles of cells that support perception and action by constantly attempting to match incoming sensory inputs with top-down expectations or predictions. This is achieved using a hierarchical generative model that aims to minimize prediction error within a bidirectional cascade of cortical processing. Such*

---

[16] Much of the literature refers to `predictive coding'. However, we do not limit ourselves to schemes designed to predict continuous variables, like the acoustic properties of a voice. Instead, we mean all forms of predictive processing, including those that deal in categorical variables like phonemes, words, and sentences, so will refer in the following to `predictive processing'. We thank Thomas Parr for this comment.
[17] Miscues are "window on the reading process" (Goodman and Burke, 1973). They uncover both the lower and higher level processes readers undertake as they read (decoding phonological and graphic information, as well as predicting. sampling, confirming, and correcting).

*accounts offer a unifying model of perception and action, illuminate the functional role of attention, and may neatly capture the special contribution of cortical processing to adaptive success"*.

This is a hierarchical process in that it involves multiple levels (Ding *et al* 2015) and multiple timescales (Keitel et al 2018). It is facilitated firstly by downward passing of information in the cortex (Bar *et al* 2006), and secondly by feedback loops of thalamo-cortical circuitry (Alitto and Usrey 2003, Briggs and Usrey 2008, Kveraga *et al* 2007) shown in **Figure 4.** There is no direct link from either the visual system or the auditory system to the relevant parts of the cortex; in both cases incoming information is first sent to the thalamus for processing. Here cortical predictions generate a difference signal relative to incoming data from the optic nerve, which is then fed back to the cortex as a measure of surprisal (Parr *et al* 2018) which is used to update predictions in the cortex

This is a non-linear signal processing operation. New data comes in (as it is constantly doing), you update your current hypothesis on the basis of this incoming data through Bayes Rule, a mathematical relation which your mind automatically implements (Clark 2013). This happens subconsciously in such a way that these predictions actively and efficiently facilitate the interpretation of incoming sensory information and directly influence conscious experience (Panichello *et al* 2013). This updating implements a causal loop that makes the process non-linear (**Figure 3**).

We often fill in what we think is right (based in previous experience) even if it's not what is actually there. A good non-technical presentation is Yon (2019). Through these processes, vision works in a gestalt or holistic way[18] (Kandel 2012, Kandel 2016), whereby one rapidly sees the whole. As explained by Orbán *et al* (2008), humans extract chunks from complex visual patterns by generating accurate yet economical representations and not by encoding the full correlational structure of the input. Thus our brains do not have to notice the parts first in order to construct the whole, rather the whole is perceived first and the parts are usually perceived later. We have all experienced how in a new environment, we tend to notice the big picture first: we see the general outline of things before we start taking account of the details. So babies and young children who are still learning what things are also do this. Babies consciously recognizing their mother for the first time take in and respond to the whole face and the eyes, they get to know the other parts gradually later. A toddler first sees a dog or cat in its entirety, they don't have to first identify and learn the parts of the animal before they can assemble them into a whole. A plastic doll with movable parts is not perceived as a doll only once the child has learned it is made up of arms, legs, a head with many strands of hair, and a torso. A real or toy car is understood first as a whole, not by building it up from wheel to windscreen wiper, indeed in general one does not know (or need to know) what all the parts are.

**A physical action aspect** In many cases, this process involves physical action: the nature of the world is

---

[18] For up-to-date views on gestalt psychology in perception, see Wagemans *et al* (2012), Wagemans *et al* (2012a), Isaac and Ward (2019).

tested by acting on it and seeing if the outcomes are as predicted (Friston *et al* 2017). There is a cycle:

$$\text{predict} \longrightarrow \text{perceive} \longrightarrow \text{act} \longrightarrow \text{predict (repeat)} \qquad (2)$$

where the boundary between the brain and the world can be characterised as a Markov Blanket (Friston 2003, 2010). Parr *et al* (2019) state that the variational perspective of cognition formalizes the notion of perception as hypothesis testing, and treats actions as experiments that are designed partly to gather evidence for or against alternative hypotheses. Thus expectations come from experience (Yon 2019). In fact
> "*Brains construct hypotheses and test them by acting and sensing…Brains sample information, hold it briefly, construct meaning, and then discard the information*" (Freeman 2002).

**Social Context**: All of this takes place in social contexts, and constitutes socio-cultural and linguistic practices involving social engagement with role modelled behaviour (Longres 1990) leading to a social Bayesian brain (Otten *et al* 2016). Intentions and meanings of others drive the understanding of implied features and linkages of a text (Donald 2001, Frith 2007, Friston and Frith 2015).

**2.2     Listening and reading are forms of perception**

Speech can be regarded as a form of perception; it is a predictive correction process based on prior knowledge (Sohoglu *et al* 2012). Written language is perceived in this way too. So if a young child has a word pointed out to her and is told that this says `cat' or `giraffe' or `Granny', she will perceive the entire word, just as she perceives an entire toy doll, car, or train. Gestalt imagery is a critical factor in language comprehension (Bell 1991). This can initially happen before a child understands the alphabetic principle. It depends on experience and context, and is why very young children are sometimes able to read brand names such as McDonalds, Coca Cola or KFC – they are seeking the meaning of the writing they encounter in its context, and are reading the sign as a whole (Harste *et al* 1984, Bua Lit 2018).

     For competent readers, reading is fundamentally a contextual, holistic process. Sense making of words and sentences occurs: they are generically understood through contextual dependence on meaning, rather than by stringing together the component parts to reach a cumulative point of comprehension. So contextual word recognition occurs. In English and in other languages, it is common for words to have meanings and pronunciations that are contextually dependent such as "wound", "wind": She wound the clock, his wound hurt; wind the clock, the wind is blowing hard, he planned to wind up his opponent. Reading always involves filling in implied contextual information on the basis of prior experiences and cultural expectations. This happens both at a local level (Who is "she?", "What hurt him?", " Why did the clock stop?" and so on) and at a more global level (Does mention of an owl imply bad luck or wisdom? What does the phrase "The Holocaust" mean?, etc). This is a key part of understanding when reading (Donald 2001; Box 1 in Castles *et al* 2018), see **Figure 3**. The brain subconsciously corrects errors and fills in missing words through the predictive processing process. This is the major reason that proofreading a text

you have written is so difficult: you literally don't see what is there, you see what ought to be there because that is what your brain expects to see.

Decoding words 'accurately' with phonics rules (Shaywitz 2003) has extremely limited application in languages with opaque orthographies like English: "It is tough having a thought that sounds off colour", there is often a silent "e" as in "eye", "bye", "were", "queue", "quite", and so on (Strauss 2004, Strauss and Altwerger 2007). Decoding in languages with transparent orthographies is potentially easier to do from a memory perspective, their spelling being more regular and predictable (Goswami 2008). This does not however, detract at all from the predictive nature of the reading process. Indeed, as Seidenberg states, the research shows that "*there is no free orthographic lunch*" Seidenberg (2013) and that "*…there is little evidence that precocious knowledge of spelling-sound correspondences confers a comprehension advantage or that the irregularities in written English present an especial burden*" (ibid).

As mentioned above, in many cases this process involves action. Talking and listening are conjoint processes learnt together by an infant as the sounds he hears and makes move from immature babbles to conventional speech (unless deaf, where a range of other cues lead to signing). Alongside this, writing and reading what is written are conjoint processes of active perception which allows movement from immature attempts to ever better approximations to mature reading and writing (Bissex 1980, Ferreiro and Teberosky 1982, Bloch 1997). Learning to read and comprehend can happen without learning to write; it is however not possible to learn to write without reading. When learning to read and write in ways based in integrated understandings which centre on purposeful uses of print, attention is on meaning as texts are written and read in concert. These processes reinforce and support each other symbiotically.

**2.3    Critiquing the neuroscientific basis for the current reading orthodoxy**

The view outlined in Section 2.2 differs from current reading orthodoxy, influenced by the neuroscience work done by Shaywitz (2003) studying the brains of children with problems learning to read. This view is strongly represented by the writings of Helen Abadzi. She states,

> "*To read and make sense of a text, our brains must first link together lines perceived by our eye receptors. The visual areas of the brain register these individual features, and, with practice, they combine them into the letter shapes used in various cultures*" (Abadzi 2017:4).

But she then states as regards mathematics,

> "*...we group and automatize Arabic numerals. Thus, we see the number 2,365,678 not as a mere sequence of numbers but as chunks in a group that gives a sense of magnitude. Similarly we assemble letters and numbers into complex mathematical equations… And how does meaning arise from these grouped shapes? The brain interprets them according to needs in the environment*" (Abadzi 2017:4).

This is correct. She does not however draw the corollary that the same thing happens in reading text. In general, as in the case of mathematics, the cortex chunks the text and interprets it on the basis of environmental context, seeing whole words and phrases rather than strings of letters.

Abadzi makes the following statement "... *The neuronal pathways originate from the visual cortex and move forward, linking sounds and subsequently linguistic processes*" (Abadzi 2017:5). This is contradicted by the studies we have mentioned above of how sensory processes work. Contrary to her view, prediction and filling in takes place both between cortical layers (Bar *et al* 2006, Rauss and Pourtois 2013) and via thalamo-cortical pathways (Alitto and Usrey 2003, see **Figure 4**) whereby downward feedback signals affect what one sees and hears. They are omitted from Abadzi's Figure 1 (Abadzi 2017).

The process is not a one-way process from sensory organs to the cortical layers, and it is not a one-way process from incoming sensory data to output. That is a basic misrepresentation of how the brain actually works. Curiously she states in the next paragraph "*The evidence points to a hierarchical, cascaded, interactive model of word recognition, in which top-down feedback consolidates fast feed-forward influences via recurrent processing loops*"(Abadzi 2017:5). Indeed so. This is what underlies the real reading process. This correct statement contradicts her previous one.

She then goes on to say, "*Thus, reading involves closely timed sequences, where performance at each stage must be optimized to give reliable and timely input to the next. The meaning-related areas are at the end of this path. It is necessary to lift the print off the page before interpreting a text*" (Abadzi 2017:5). In reality (Bever 2009, Bever and Poeppel 2010, Bever 2017) we predict what will be there as we read the words on the page in any detail - in essence interpretation precedes lifting the details of print off the page. This is confirmed by detailed EEG studies (Monsalves *et al* 2014).

Abadzi states later "*Instead, ''comprehension'' is often used to signal inferences or predictions. These require more knowledge than offered in a text*" (Abadzi 2017:8). Precisely so. That is why reading is a contextual process of interpretation, extending to a psycho-linguistic guessing game in the case of complex texts (Goodman 1967, Bever 2009).

**2.4    Reading and predictive correction: jumbled words**

A famous illustration of this predictive property is on our ability to read jumbled words (Seidenberg 2017:85-99, Rayner *et al* 2006): yu cn raed this evn thogh wdrs wonrg and messd up. This is the subject of an informative comment by Matt Davis[19] and the thesis work by Rawlinson (1976). It is significant because it gets to the heart of the predictive reading process. It is summarized by Rawlinson[20] as follows:

> *"My conclusions, and these are open to question of course, were that: Letter features are processed through a route of letter classification/identification. Middle letter identification proceeds largely independently of position. Higher level units seem to be significant only for the beginnings and endings of words. Information from the middle letters may operate via a sampling/probability system (rather than absolute accuracy). That is, you can have sufficient letters, even though in the wrong position, for the brain to `recognise' the word. My end model was of a multiple access system `allowing some direct use of features without precise letter identification, use of word length*

---

[19] See http://www.mrc-cbu.cam.ac.uk/personal/matt.davis/Cmabrigde/.
[20] See http://www.mrc-cbu.cam.ac.uk/personal/matt.davis/Cmabrigde/rawlinson.html.

*information, and some structuring of phonemic or syllabic units, as well as incorporating a sampling recognition system using letters or their attributes directly.' I suggest the experiments `demonstrate the considerable flexibility of the reading process'. Stimulus sampling theories seem to apply more than simple phonetic theories of word recognition. As regards learning to read, `when the child is beginning to learn to read (s)he already has a highly refined set of skills not only for dealing with the known world but also for selecting and using information from the unknown world'. `Word recognition skills develop which are not only not taught but which develop despite sometimes fairly specific teaching in alternative skills."*

This key evidence strongly supports the predictive processing understanding of reading.

## 2.5 The centrality of social context

This predictive process is always shaped by social context (Donald 2001, Frith 2007). Friston and Frith (2015) explain that in the case of speaking and listening, communication is centred on inference about the behaviour of others: *" We are trying to infer how our sensations are caused by others, while they are trying to infer our behaviour…. This produces a reciprocal exchange of sensory signals that, formally, induces a generalised synchrony between internal (neuronal) brain states generating predictions in both agents."*
This is what many call "mindreading" (Donald 2001:59-62, Frith 2007:16, Heyes and Frith 2014). Fabry (2017) gives an account of prediction error minimization that is fully consistent with approaches to cognition that emphasize the embodied and interactive properties of cognitive processes. Constant *et al* (2019) give the predictive processing view of cognition extending beyond skulls. In short, the brain is a *social Bayesian brain* (Otten *et al* 2017): social knowledge can shape visual perception. Literacy essentially involves the same issues, and is therefore a social practice (Street 1984, Barton *et al* 2000).

## 2.6 The nature of language processing across modes

Farmer *et al* (2013) summarise how the predictive processing view extends to language processing across modes. It applies equally to spoken, written, and sign language, the latter being an important form of language where no phonemes occur. There is no divergence as to how these various language modes are handled by the brain. Berent (2020) summarises as follows:

*"Linguistic principles themselves transfer across modalities. An early exposure to sign language helps because some of its rules are relevant to the later acquisition of English. Language is neither speech nor sign, but an abstract algebraic system that can emerge in either system."*

The same applies to spoken and written language. They are all realisations of the same abstract relations (Huybregts *et al* 2016). Similarly, significant aspects of learning to read and write are transferred to learning new languages (Bialystok *et al* 2005). A key point is made by Seidenberg *et al* (2020):

*"Reading depends on speech. Students do not relearn language when they learn to read; they learn to relate the printed code to existing knowledge of spoken language. Writing systems are codes for*

> *representing spoken language. The structure of spoken words in English—the fact that they consist of sequences of phonemes, syllables, and morphemes that are associated with meaning—is reflected in their alphabetic representations. Learning about the written code is easier for students who know more about characteristics of spoken words that it represents. Individual differences in knowledge of such properties of spoken language at the start of formal instruction have an enormous impact on students' progress".*

Significant implications of this are both the value of enriched language input and of ensuring comprehensible input (Krashen 2017, Krashen and Mason 2020) for all children, with particular attention to children learning in difficult conditions and learning multilingually.

## 2.7 An integrative predictive processing view of reading

Strauss *et al* (2009) summarise the predictive processing view of reading as follows:

> *"Whereas the classical neuroanatomic view is most consistent with a bottom-up, information processing model, the emerging view supports an interactive, constructivist model. The cortex either promotes or inhibits the very input being transmitted to it from the eyes, ears, and other sensory receptors. The psychological interpretation of this neuroanatomic arrangement is that the cortex selects evidence to confirm or disconfirm its predictions. It anticipates what will be seen and heard using knowledge stored in memory. Both this new neuroanatomical view and its psychological reflection are consistent with a transactional socio- psycholinguistic model of reading. Drawing on extensive comparisons of expected and observed responses from oral reading miscue studies, this model of reading emphasizes the fundamental importance of effective and efficient prediction and confirmation in the construction of meaning."*

This holistic, meaning-construction view of reading and writing is confirmed by eye movement analysis, miscue studies, and the ability to read partly hidden or garbled text,

## 3  Emotions play a key role in underlying cognitive functioning

Emotions play a key role in underlying normal cognitive functioning from birth onwards. Innate affective systems underlie and shape all brain functioning, including communicating in speech and writing. Genetically determined inbuilt emotional systems functioning via reticular activating systems (**Figure 5)** stimulate and guide all cognition and learning from birth (Panksepp 1998, Panksepp and Biven 2012, Ellis and Solms 2017), and so play a key role in particular in oral and written language learning.

### 3.1 The key role played by emotions in normal cognitive functioning

A key factor in all brain function is the emotional systems that underlie motivation in life in general (Panksepp 1998, Damasio 1999, 2000, Panksepp and Biven 2012, Ellis and Solms 2017), and in particular for children in the classroom (Willis 2006). They are also key in language development (Greenspan and

Shanker 2004: 210). Railton (2017) states

> *"Recent decades have witnessed a sea change in thinking about emotion, which has gone from being seen as a disruptive force in human thought and action to being seen as an important source of situation- and goal-relevant information and evaluation, continuous with perception and cognition.… The affect and reward system—affective system, for short— is the central locus of the learning processes, evaluative representations, and spatial mapping and simulation essential for the reasons-sensitive action guidance."*

An important feature is that all memories have an emotional tag, either positive or negative.

Because of their great significance for learning, we will discuss the affective systems in more detail in the next section. A crucial distinction exists between the primary (genetically determined) affective systems and associated emotions, and the secondary (socially determined) emotions.

### 3.2 The primary emotional systems

Innate affective systems (Panksepp 1998, Davis and Montag 2019) are 'hardwired emotional systems' that all babies are born with. They underlie and shape all brain functioning, and result in felt emotions. These are our evolutionary inheritance, genetically determined to be what they are because they were essential for our survival in the distant past (Panksepp and Biven 2012, Ellis and Solms 2017). They are also the initial and ongoing propensities which all babies and young children bring to any learning.

These primary emotional systems function via the ascending reticular activating system: diffuse projections to the neocortex from nuclei in the arousal system (roughly: the limbic system) that spread neuromodulators such as dopamine and serotonin to the cortex. A particular example (the SEEKING system) is shown in **Figure 5**. These primary affective systems both affect immediate behaviour, and underlie brain plasticity by shaping neural connections because they form the "value system" for Gerald Edelman's Neural Darwinism (Edelman 1987, Ellis and Toronchuk 2005) whereby neural network weights are affected by experience. Panksepp (1998) lists seven such primary emotional systems; Ellis and Toronchuk (2013) suggest a further two, agreeing with claims by Stevens and Price (2015). We will now briefly review those that are most important for early learning.

**A) The search for meaning** A core feature of psychology is the search for meaning (Frankl 1985). This drive is associated with the "SEEKING" system (Panksepp 1998) which is the primary hardwired emotional system all babies are born with. It is a prime motivator for all they do: exploring the world around and trying to understand it so that it becomes predictable (and this is what the Predictive Processing model is about). In particular they want to understand the meaning of what their primary caregiver does (Greenspan and Shanker 2004). The SEEKING system and the search for meaning play a key role in all cognitive learning, and in particular learning to speak, because our brains are wired to search for meaning and intention (Frith 2007), and it is language that enables the joint construction of meaning (Evan 2015). This leads us to question what happens to young children's impetus to learn when, on entering formal education, they are expected to set aside their expectation (active since birth in informal settings) that seeking and making meaning drives learning, and replace this with working out how to give the teacher what she asks for, irrespective of the sense it makes to them[21].

**B) The need for community and belonging** The second core primordial emotional need is that of belonging to a community (Stevens and Price 2015) because we have a social brain (Dunbar 1998). In the case of babies and young children, Panksepp labels this the PANIC/DISTRESS system, which has to do with the strong need to be in the secure presence of the primary caregiver, and the panic and distress experienced when this support is removed (Panksepp 1998). In the broad context of society, it should more properly be labelled the BELONGING/AFFILIATION system, which includes both mother/child bonding and the deep need to belong to social groups (Ellis and Toronchuk 2013, Stevens and Price 2015).

This interaction between mother and child involving the development of relationship, facilitates the emergence of spoken language in the child. The importance for language development of the emotional need to interact intensely with the primary caregiver is explained clearly in *The First Idea* (Greenspan and Shanker 2004). Such intense interaction provides rich stimulus for language use in purposeful contexts, contrary to Chomsky's claims of lack of sufficient stimulus to enable language learning. The key contextual feature in early childhood, shaping this all, is the relationship with the caregiver. Tomasello states:

> *"The glue that holds this all of these factors together is always the child's attempts to understand the communicative intentions of other persons as she interacts with them socially and linguistically ..... children learn words most readily in situations in which it is easiest to read the adult's communicative intentions ..... usage based linguistics holds that the essence of language is its symbolic dimension, that is, the ways in which human beings use conventional linguistic symbols for purposes of interpersonal communication"*(Tomasello 2003:44,49,283).

The kind of informal learning which is stimulated through this need for community and belonging is determined and shaped by situated cultural practices used and valued in particular environments (Rogoff *et al* 2016). This suggests the strong case for encouraging and enabling learning written language in similar ways, as discussed in depth in Ferreiro and Teberosky (1982), and demonstrated in Bissex's *Gnys at Wrk: A Child Learns to Write and Read* (1980) and *Chloe's Story* (Bloch 1997). We return to this in **Section 6**.

**C) The role of play** Play is one of the primary emotional systems (Panksepp 1998, Ellis and Toronchuk 2013, Ellis and Solms 2017), leading to rough and tumble play in all mammals, and to various forms of play, including imaginative/ symbolic play, in humans. Play, which evolved tens of millions of years before language, has great significance for learning (Gray 2017:120-122). Gray states that the varieties of play

---
[21] This is not to imply that informal learning does not involve working out what the mother or other wants, but that informal learning has a strong self-motivated voluntary aspect.

match the requirements of human existence.[22] As stated by Boyd (2018:19),

> *"[It] offers a way of learning species-typical skills by detaching them from serious mode, testing them in safe circumstances in exuberant fashion so that trial and error can refine them at low risk. Play has been so beneficial in the young of so many species that it has evolved to become self-motivating, irresistible – sheer fun".*

It involves children in symbolic thinking, exploring, and discovering alternative options and their outcomes, and hence leads to creative thinking and understanding (Bruce 1991).

Behaving symbolically (Deacon 1998) as children do in pretend/imaginative play, underpins literacy learning, a 2nd order symbolic system (Vygotsky 1978, Stone and Burriss 2016). This imaginative/symbolic play arising from the PLAY system has fundamental and ongoing relevance from babyhood onwards: early word play and action play using songs, rhyme, and alliteration (Bryant *et al* 1990) are all practiced voluntarily by toddlers and young children as they develop a feel for the repetitions and rhythms of their languages. Such behaviour contributes to learning to read, especially when bridging connections can be made from oral to written forms, for instance with rhymes. Children come to sense the 'tune on the page' (Meek 1988) by encountering these oral wordplays in print, with illustrations to provide initial clues to meaning. Moreover, intrinsically motivated, self-directed child exploration and discovery of written language through play (Bruce 2015) and story (Gussin Paley 1990) connected to children's current concerns and interests, leads to deep engagement (Cooper 2009, Roskos *et al* 2003, Roskos and Christie 2011).

### 3.3 Emotions, play and stories

Language processing involves salience and attention in accord with the predictive processing paradigm (Zarcone *et al* 2016). Reading and writing in authentic contexts involves conveying and negotiating meaning, facts, stories, and emotions between authors and readers (Meek 1988).

Play is described as story in action by Gussin Paley (1990). This reveals the significance of stories for early learning in both spoken and written language (Nicolopoulou *et al* 2015). As storytelling animals we make sense of our lives through stories (Gottschalk 2012): it is a powerful form of meaning making and social sharing (Redhead and Dunbar 2013), and we feel compelled to share our stories, factual and fictitious, with one another. Children's attention, imaginations and thinking are activated when immersed in formal or informal contexts in stories (Stanley 2012) - life stories, history of families and communities, or imaginative stories. Authentic language use and learning through stories (Egan 1989, Sugiyama 2017) offers adults and young children power and voice. Encouraging children to tell and compose their own stories, and valuing these, makes important connections to children's home funds of knowledge and identity (Moll *et al* 1992, Esteban-Guitart and Moll 2013). Prediction, emotion, and the embodied mind are fruitfully entangled together in these contexts (Miller and Clark 2018).

---

[22] https://www.psychologytoday.com/za/blog/freedom-learn/200810/the-varieties-play-match-requirements-human-existence?fbclid=IwAR1sy4sMpHrzpimt4Yu_o9du27kR5-1sYaZelukRR2wSU_Q1JRuOeWJk0A0.

**3.4 The secondary emotions: extrinsic motivation**

Secondary (social) emotions such as pride and shame are also important in mental life. They are not genetically determined due to evolutionary processes, as the primary emotions are. This is because unlike the primary emotions, there are no associated ascending systems in the brain. They are socially determined as a result of social processes and play an important role in shaping socio-cultural interactions. They piggyback off the BELONGING/AFFILIATION system which underlies socialization.

Extrinsic rewards tend to be used very early in school through marks, stars, competitions, prizes, and so on. The emotional outcome can be both positive (praise, high marks) and negative (tests failed, low marks). Affirmation is indeed a strong motivator that leads to positive behavioural outcomes, but overly competitive or punitive aspects can have either positive or negative behavioural outcomes: they may result in greater effort, but they may also result in humiliation, anger, despair, and demotivation.

3.5 **Emotion, reading, and literacy learning**

We do not necessarily consciously acknowledge this, but negative emotional tags are one of the most serious stumbling blocks to learning. This is a well-established fact in the case of mathematics education (Carey *et al* 2017). In the case of reading assessments, the Early Grade Reading Assessment (EGRA), has been developed for wide use in the Global South. It is based on the Diagnostic Interpretation of Basic Early Literacy Skills (DIBELS) in the USA, which has been criticized for the emotional upset it causes some young children (Goodman 2006). Once demotivated, it is very difficult for children to succeed.

We need to pay much more attention to this as it is a potentially critical factor in literacy learning problems in classroom contexts (Meyer and Turner 2006, Immordino-Yang *et al* 2019). However emotional or affective systems are not mentioned by Shaywitz (2003), Dehaene (2010), Abadzi (2017), or Castles *et al* (2018), although the latter mentions the closely associated features of boredom (p.14) and motivation (p.26). Emotion is however mentioned by Hruby and Goswami (2011). We regard the emotive aspect of learning to read and write as critical to creating proficient readers, starting at the earliest ages via the caregiver/infant interaction (Greenspan and Shanker 2004),and illustrated when young children learn to write and read together as in a six year biliteracy project (Bloch and Alexander 2003 pp104-114)  and a small home based early literacy project (Alexander and Bloch 2010, pp 204-210) carried out by one of us with colleagues; this was as much the case with isiXhosa as it is with English only (Bloch 1997).

The great importance of motivation for learning to read and write (Wigfield et al 2016) can be understood as arising from and developing  these primary and secondary emotional systems.

# 4    There is not the fundamental difference between oral and written language that is often alleged

The claim is made frequently by many involved in literacy education that it is an evolutionary fact that oral language, i.e. listening and speaking, represents the only 'natural language', acquired in social contexts without teaching (Shaywitz 2003). Written language, i.e. writing and reading, is understood to be a cultural and artificial invention needing specifically structured teaching, with components initially simplified and taught separately (Wolf 2008, van Rooy and Pretorius 2013, Spaull and Pretorius 2019:5).

This claim comes in many forms. Gough and Hillinger (1980) describe reading as an "*unnatural act*". It is captured in the following statement by Willis:

> *" Reading is not a natural part of human development. Unlike spoken language reading does not follow from observation and imitation of other people"* (Willis 2008:2).

This remark illustrates the contradictory nub at the heart of this debate: the author holds this foundational view, which leads her to to believe in the necessity for young children to 'crack the code' in decontextualized ways. Like some others who hold authority as neuroscience experts (e.g. Wolf 2008), once this is achieved, she reverts to a meaning based understanding and approach.

Wolf (2018) states new neural circuitry was necessary for reading because reading is neither natural nor innate; rather, it is an unnatural cultural invention that has been scarcely 6,000 years in existence. By contrast, we view both oral and written language are equally 'natural' (Goodman and Goodman 2013)**.** This is because both are social constructs, developed in evolutionary terms as successive modes of symbolic communication, the latter piggybacking on the former, when the need arose as part of human development. They can both be learnt by essentially the same social processes (**Section 6.2**), with the symbolism of thought realised in different ways (oral and written); and the same is true for sign language and Braille.

## 4.1 Naturalness of oral and written language: an innate language system?

Shaywitz (2003:45; 49-50) states,

> "*Reading is more difficult than speaking…. Spoken language is innate. It is instinctive. Language does not have to be taught. All that is necessary is for humans to be exposed to their mother tongue. Although both speaking and reading rely on the same particle, the phoneme, there is a fundamental difference: speaking is natural and reading is not. Herein lies the difficulty. Reading is an acquired act, an invention of man that must be learned at a conscious level. And it is the very naturalness of speaking that makes reading so hard."*.

She justifies her views of the difference between reading and writing as follows (Shaywitz 2003:50*):*

> *"Profound differences distinguish reading from speaking … Reading is not built into our brains. There is no reading module wired into the human brain. In order for children to read, man has to take advantage of what nature has provided: a biological module for language*".

That is, she is claiming the key to the difference between oral and written language is innate properties and how they underlie brain development.

Shaywitz is relying on Chomsky's idea (Chomsky 1965, 1975) of an innate language module in the brain: a *Language Acquisition Device* (LAD). But there are in fact no innate cortical modules in the brain representing evolutionary-based hard-wired knowledge of any kind; this is not possible for evolutionary, developmental, information theoretic, and physiological reasons (Ellis and Solms 2017). Rather we are provided with brains that are highly plastic and able to adaptively learn through ongoing experience with the physical, ecological, and social environment. We have learning-ready brains.

What is preset is the primary emotional systems that guide action (**Section 3.2**). But above all, Shaywitz fails to recognize that the process of learning to listen and speak is just as much a learning process as is learning to read and write. The implication of this view is that such 'natural', oral language is acquired effortlessly. We question and contest this. Babies cannot talk when they are born. They learn through a complex, extensive and persistent process involving social interactions, during the first few years of life, as stated by Kuhl:

> " The learning processes that infants employ when learning from exposure to language are complex and multi-modal, but also child's play in that it grows out of infants' heightened attention to items and events in the natural world: the faces, actions, and voices of other people" (Kuhl 2010:716).

And the same is true for written language. Both have to be learned at a conscious level and both involve teaching. This is discussed in **Section 6**.

### 4.2 Language readiness versus a Language Acquisition Device

There is no LAD as envisaged by Chomsky on behavioural grounds. As Evans (2020) states,

> *"Everyone agrees that our species exhibits a clear biological preparedness for language… What is in dispute is the claim that knowledge of language itself – the language software – is something that each human child is born with. .. a 'language organ' … containing a blueprint for all the possible sets of grammar rules in all the world's languages."*

Pinker (2003) called this a `language instinct'. The problem is that Chomsky proposed his LAD without taking into account the biological processes whereby the brain comes into being. If you bring biological reality into the picture by considering this, such a LAD cannot exist for developmental, genetic, and evolutionary reasons, as explained in depth in Ellis and Solms (2017). We summarise the main reasons thus:

First, there is no way that the precise details of the billions of neural connections in the neocortex can be guided by developmental processes: the refined detailed nature of the connections make that impossible. Rather the detailed synaptic connections are initially made randomly, and then refined on the basis of experience (Wolpert *et al* 2002). They are not directly genetically determined.

Second, there is not a fraction of the genetic information available in the human genome needed to shape such detailed neuronal connections. It contains about 30,000 genes, which are needed to code for the entire body: heart, lungs, liver, digestive system, skeleton, skin, *etc.,* and in particular to set up the large scale brain structure. There simply are not enough genes to determine the detailed cortical structure with billions of connections. In any case only a fraction of those genes are specifically human genes that can conceivably be associated with grammar.

Third, setting aside these two critical issues, it is not remotely plausible that the kind of detailed grammatical structures investigated by Chomsky would have been of such a vital importance that they would have resulted in evolutionary selection because they affect survival probabilities so crucially. Selection for an overall language capacity, yes that is critical: but for this kind of detailed grammatical structuring, no way. Because of the predictive processing nature of language perception (Section 2), minor grammatical errors do not harm understanding of the message being conveyed and are not needed for survival. As discussed above, the brain automatically makes the needed corrections.

These considerations are decisive (Ellis and Solms 2017): there is no genetically determined LAD. The real situation is that we possess a language ready brain with a generic symbolic capacity (Deacon 2003) which in suitable social contexts learns to understand both spoken and written language, or sign language in the case of deaf people. Evans (2014, 2020) develops this all in a clear way, emphasizing how as more data has been collected, the claims of grammatical universals have weakened over time.

There are however two further arguments to consider: Chomsky's Poverty of Stimulus argument, and the issue of where language universals come from.

**4.3     The Poverty of Stimulus argument**

There are three counters to this claim made by Chomsky that there is not sufficient evidence provided to children for them to be able to learn the grammatical rules of their home language as a social process.

First, as pointed out by Lewis and Elman (2001), Chomsky's poverty of stimulus argument(1975) fails to hold once stochastic information is admitted. The properties of language in question is shown by them to be learnable with a statistical learning algorithm. They show that simple recurrent networks are able to provide the correct generalizations from the statistical structure of the data. Pullum and Scholz (2002) detail how the linguistic nativist position noted above is not supported by the data. Amodei *et al* (2016) show how statistical learning can be done in practice via an end-to-end deep learning approach. This is in line with the predictive processing view. Friston *et al* (2020) propose that the neuronal correlates of language processing and functional brain architectures should emerge naturally, given the right kind of generative model. The basic issue is that language processing is not in fact rule based, it is based in statistical correlations (see **Section 5**).

Second, is there in fact a poverty of stimulus? We claim there is not in normal situations, where massive stimulus is provided by the main caregivers, as emphasized by Greenspan and Shanker (2004). Rogoff (2003:69) describes human beings as 'biologically cultural' and states,

> *"Whether or not they regard themselves as explicitly teaching young children, caregivers routinely model mature performance during joint endeavours, adjust their interaction and structure children's environments and activities in ways that support local forms of learning"*.

The stimulus which occurs for language learning crucially involves the strong emotional link discussed in **Section 3**, as well as continuous demonstrations of (culturally) conventional or mature speech in action, to which children gradually adjust their immature speech attempts. These are the basis for statistical learning processes.

Third, this ability to learn either spoken or written language through such interactions is significantly strengthened when these interactions are laden with positive affect, as discussed in **Section 3**. This enhances the motivation to transact with and understand the message being conveyed, and hence also to grasp the grammatical patterns by which it is conveyed.

## 4.4 Language universals

Where then do language universals come from? A plausible view is that they are due to essential syntactic limitations that must necessarily apply to any language whatever due to the requirement that it be an adequate symbolic system for representing the world around. They arise due to fundamental semiotic constraints on any symbolic representation of our experiences and environment, as explained in detail by Terrence Deacon :

> "*Many of these core language universals reflect semiotic constraints, inherent in the requirements for producing symbolic reference itself… these constraints shape the self- organisation and evolution of communication in a social context. .. combinations of words inherit constraints from the lower order mediating relationships that give words their freedom of mapping. These classes of constraints limit the classes of referentially consistent higher order symbol constructions*" (Deacon 2003, pp. 112, 118).

That is, they arise because language must provide a meaningful representation of the world around us in order to be useful. Tomasello reinforces this view (Tomasello 2003:18).

## 4.5 Naturalness and evolution of reading and writing

We have asserted that there is not the fundamental difference between listening/speaking and reading/writing that is often claimed on the basis of evolutionary arguments and the alleged existence of a LAD in the brain. Oral and written language are both social practices driven by the communication imperative of the social brain (Dunbar 1998), learnt through socio-cultural processes. They both evolved in similar ways through the social processes of cultural evolution.

An examination of the historical record will show that oral language first evolved as a crucial cultural invention between 70,000 and 30,000 years ago (Harari 2011:23-28), and writing evolved as a second cultural invention piggybacking on the first between 3,500 and 3,000 BC (Harari 2011:137-148). Neither is hardwired in the brain, as discussed above; both are socially transmitted down the generations. Spoken language evolved to enable efficient human bonding (Dunbar 1993) in particular enabling communication among kin (Fitch 2005); Tomasello (2000) and Donald (2001) give broadly consistent viewpoints. Writing later evolved to solve the problem of cooperation in large groups by transcending the severe limitations of our evolved psychology through the elaboration of four cooperative tools – (1) reciprocal behaviours, (2) reputation formation and maintenance, (3) social norms and norm enforcement, and (4) group identity and empathy (Mullins *et al* 2013). As a major extension of oral language, writing evolved to allow communication over space and time and record keeping over time in unparalleled ways.

While there can be contestation about the details, the fundamental issue is clear: both oral and written language evolved in broadly similar ways to enable the cultural evolution of human cooperation (Fitch 2005). Thus we suggest that the statement "reading is unnatural" could usefully be replaced by

> *"Reading and writing are both cultural practices, and culture is natural."*

This proposal is strengthened if one looks at the case of sign language (Trettenbrein *et al* 2021). This is obviously also a cultural invention, and does not involve phonemes as it is a communication means for deaf people. By looking at the brain areas involved in sign language, the authors show that the human brain evolved a lateralized language network with a supramodal hub in Broca's area which computes linguistic information independent of speech. It can be realised in sound, writing, or sign.

## 5       Natural brain operations: statistical correlations and predictions, not logical rules

The foundational issue is what is the natural mode of cortical function. This underlies a question: what really is the nature of linguistics? In an important paper, Seidenberg *et al* (2020) raise this after considering the problems underlying a rule-based view of language .

### 5.1 Linguistics: Rules and Exceptions

Seidenberg *et al* (2020) summarise the ***dual-route theory of reading*** as follows: it consists of
- Rules to produce patterns such as *save-pave-gave*, which are used in sounding out unfamiliar words (or, in research studies, pseudo-words such as *mave*),
- A list of "exception" or "sight"words whose pronunciations violate the rules (e.g., *have, said, bear*) and must be memorized.

They state, "*The instructional implications of the theory are straightforward: teach children the rules (or enough to allow them to "break the code"), and help them memorize the exceptions.*" But they then ask, "*What are the rules for pronouncing written English?*", and conclude "*No one knows*". The key problem is that the dual route model does not provide a meta-rule for determining when the standard patterns apply; and without that, you cannot reliably apply the rules. Examples in English are the well-known problems in pronouncing "ou" (wound, sound, cough, tough, ought) and the problem of silent "e" (Strauss 2004), which already afflicts the standard patterns cited by Seidenberg *et al*. They then ask the key question:

> "*What if it is difficult to state the rules and how they are learned and decide on the sight words because the system **isn't rule-governed**? What if 200 years of phonics instruction has been based on a false dichotomy?*"

That is exactly the right question to ask.

## 5.2 Connectionist models: The functioning of neural networks, and language learning

Seidenberg *et al* then propose using connectionist models of the brain as providing the basis of speech and reading. Such models (Buckner and Garson 2019) are not based in following logical rules but in learning and generalising the statistics of presented text, which trains the weights of the neural network.

This is set out in Seidenberg and McClelland (1989), Seidenberg (2005), Plaut (2005), Bybee and McClelland (2005). This has to be correct, because the brain is in fact a vastly complex neural network (Nicholls *et al* 2001) with memory enabled by neural plasticity allowing statistical pattern recognition and active prediction (Carpenter and Grossberg 1991, Bishop 1995, Churchland and Sejnowski 1999, Rolls 2016) resulting in the brain in effect employing Bayes' rule at a psychological level (Hohwy 2013). The brain does not in neural terms implement a strict set of logical rules such as occur in computer programs as envisaged by Turing and von Neumann. Rather it is a Bayesian brain (Friston 2013, Seth 2014, Otten *et al* 2017) that learns statistical associations such as collocations and colligations (Hoey 2005) underlying active perception. They are developed from embodied experience as ways of conveniently describing those experiences symbolically (Feldman 2008), often in effect using metaphor as mental models (Lakoff and Johnson 1980), later generalised to abstract thought and logic.

Note that learning these statistical patterns is not the same as a learning a set of rigorous logical rules such as grammatical rules as envisaged by Chomsky (1965, 1975), which in the end are the source of the alleged problem. Rather, statistical dependencies are learnt by experience - repeated presentation of many examples – and these then form the foundation of prediction of what is to be expected, which are then used in the predictive processing way discussed in Section 2. In particular, the processing mechanisms involved in the visual recognition of novel words occur through the visual system capturing statistical regularities in the visual environment (Vidal et al 2021). Their relation to the Parallel Distributed Processing (PDP) of reading is examined in Laszlo and Plaut (2012).

Thus, while we can indeed think in a logical rule-based way (how this can occur on the basis of neural networks is discussed by Marcus (2003)), this is not the brains natural way of functioning. Our brain is a connectionist Bayesian brain whose natural mode of operation is statistical pattern learning and prediction.

As a result, the pattern-matching way of reading presented by Seidenberg *et al* (2020), summarised in their Figure 1, is exactly right. Children pick up the structure of grammar by statistical learning. They conclude,

> *"Readers do not pronounce words by explicitly applying rules; doing so would be a conscious, slow effortful process (the opposite of "fluent"). Teaching phonics by teaching rules and memorizing exceptions leaves out the statistical patterns that permeate the system and drive the fast, implicit learning process."*

This results in the "rules" being applied when they are valid, but avoids the problem of trying to determine when they apply and when they do not.

What if the language is an agglutinating language such as isiXhosa, if it does indeed have a highly regular structure? Our comment is that in this case (unlike English) it may indeed be possible to describe the language adequately via a rather strict set of rule. That will not however, change the natural way the brain operates, as just outlined. It will make it possible to efficiently learnt that language in a rule based way, because the brain's statistical predictions will be well-correlated with the outcome of that set of rules, but this will not imply that that is the best way to do so. Furthermore it should be noted that it is a matter of fact that English is the dominant world language in terms of commerce and science, and hence access to the modern economy is greatly increased by being fluent in English – where phonics "rules" are highly fallible (Strauss 2004). The problems pointed out by Seidenberg *et al* will arise when children who operate multilingually try to learn English.

### 5.3 The Development of rule based understandings in individual lives

An interesting issue that arises from this discussion is that, given that rule-based logic is not the natural mode of operation of the brain, how does it arise in developmental terms? A plausible answer is that it arises through taking part in human cultural activities of singing and games.

Music has hidden rules, embodied in the structure of rhythm: this leads to an expectation of what will come next (Huron 2008), which is in essence a rule played out over time. Children make up verse that involves rhythm (Chukovsky 1968:61,87). All play and games involve rules and an expectation they will be obeyed, conveyed in the statement "*I'm not going to play with him: he cheats*" (Elkind 2007:119). Vygotsky confirms this by saying (Vygotsky 1978:94) "*There is no such thing as play without rules*". This applies equally across cultural communities and to all kinds of games.

So our hypothesis is that the connectionist brain learns the basis of rule-driven thought through partaking in songs, rhyme, poetry, and games of many kinds. Once that understanding has taken root, it can be developed in terms of logic and then mathematics and science.

### 5.4 Formal Linguistic Theories embodying this viewpoint

The statistical associations of language occur as collocations and colligations which allow lexical priming (Biber *at al* 2006, Hoey 2005) and so underpin predictive understanding of text. How this functionality arises through embodied experience is detailed by Feldman though his neural theory of language (Feldman 2008), The outcome can be formalised in terms of *Systemic Functional Linguistics* (Halliday 1977, 1993, 2003). These alternative views of the nature of linguistics are summarised by Peter Fries in Ellis and Solms (2017), pp.125-133, based on the work of Feldman, Halliday, Hoey, and others.

The key outcome for this paper is that the rule-based view of linguistics espoused by Chomsky is not the only game in town. The other approaches briefly mentioned here are far closer to what is validated by biological reality, and has been formalised in alternative views of the nature of linguistics.

## 6   The neural route to reading with no explicit decoding can be activated from the earliest years

Language includes listening, speaking, signing, reading, and writing. Because oral and written language both evolved, it is not a coincidence that there are important similarities in the way each of them function to make and convey meaning. Both receptive aspects of language (listening and reading) and productive ones (speaking and writing) are non-linear, neurolinguistic-psycho-social processes of understanding, shaped by current knowledge and context. The previous sections related to how the cortico-thalamic circuits helped underlie the way the brain predictively searches for meaning, and the innate emotional systems that power that search. This section looks at aspects of how the cortex enables the link between writing and meaning.

### 6.1   Oral language: Meaning making in context

The first and foremost point about oral language is

> **LAN(o):** *Through speech, patterned sounds convey information, meaning, and emotion*.

This enables complex communication in socio-cultural contexts, where listening and speaking is a joint socially based interaction involving shared attention, prediction, and modelling other people's minds (see for example Frith 2007, Heyes and Frith 2014).

**The basic problem** is how we understand a linear stream of symbols representing a hierarchical structure. We have to flatten the hierarchical structure into a linear structure.

> Thus, "*Sentences are externally serial (i.e., "horizontal"): derivations are internally hierarchical, (i.e., "vertical"). That is, the computational domain of a derivation can embrace entire clauses and sentences, while the immediate processing appears to be one word after another.*" (Bever 2013). We have to learn how to handle this for both oral and written language, where the issue is the same. In the case of oral language, Bever (2017) states it thus:

> *"A sentence in everyday use combines a stream of sound, with rhythm and pitch variations, with memorized units of meaning, an organizing structure that recombines those meaning units into a transcendental unified meaning that includes informational representations, general connotations, and specific pragmatic implications unique to the conversational context. In other words, each sentence is a miniature opera of nature."*

Ding *et al* (2016) explain that in speech, hierarchical linguistic structures do not have boundaries that are clearly defined by acoustic cues and must therefore be internally and incrementally constructed during comprehension. This is the predictive processing process that underlies listening to speech.

Cortical activity at different timescales concurrently tracks the time course of abstract linguistic structures at different hierarchical levels, such as words, phrases, and sentences. This is how the brain handles the problem flagged by Castles *et al* (2018):

> ``*The segmentation of an acoustic signal does not correspond in any straightforward way with segmentation at the phoneme level: In continuous speech, phonemes overlap and run together*".

From a larger perspective, understanding speech involves a 'psycholinguistic guessing game' such as is characterised by Goodman (1967), Tovey (1976), Flurkey *et al* (2008) (Bever 2009) in the case of reading. It usually has a major social component (What does this refer to? Where did that take place? Why are they saying this? Is there a hidden agenda? and so on). The predictive processing underpinnings of this process are explained by Friston and Frith (2015). These enable the process of ``mind-reading'' mentioned earlier: a key social skill leading to a theory of mind (Conte *et al* 2019).

**6.2  Written Language: Meaning making in context**

The first and foremost point about written language, parallel to **LAN(o)** above (**Section 5.1**)**,** is

**LAN(w):** *Through written text,[23] printed symbols convey information, meaning, and emotion*.

This enables complex oral and written communicative transactions in social contexts (Vygotsky 1978, Rosenblatt 1982) across distance and time.[24]

**Predictive reading** Similarly to when processing spoken language, when reading complex texts, there is never enough information in a sentence to fully convey the intended meaning. Thus in order to read or to listen, we use prediction and then comparison with incoming data, as in the case of all other senses, and in agreement with the predictive processing model of the mind (Section 2). Competent readers do not read by assembling phonemes into words and words into phrases as Shaywitz (2003) claims. They read phrases as a whole in a way that makes sense in terms of context and making meaning overall, predicting what text will come next as they do so (Goodman 1967, Bever 2009). Not all words need to be read (**Figure 3**).

**Multiple cueing systems** Readers predict meaning using multiple cueing systems (**Figure 2**): semantic, directly involving meaning, grapho-phonic, the look and the sound of the language, and syntactic, its grammatical structure (Goodman 1967, Goodman and Burke 1973, Clay 1991, Bergeron and Bradbury-Wolff 2010). Each is drawn on as required to understand the text, even when using a language which has transparent orthography, such as Spanish or isiXhosa. This is because these cueing systems work together to support the essence of reading. We strongly suggest that when children are first taught to rely only or mainly on decoding and word level accuracy, this hinders or blocks their developing metacognitive abilities to self-monitor and self-correct for meaning using various cues (Clay 1991, Juliebö *et al* 1998). Moreover, children learning multilingually, who have become habituated to mainly attend to decoding accurately, are likely to struggle when they have to start learning to read in an additional language like English. With the combination of its opaque orthography, and their emerging understanding of the language they have a considerable challenge: attending to different cueing systems within flexible languaging practices ( Garcia and Wei 2014, Makalela 2014) would make their progress in reading with meaning far more likely.

---

[23] And their extensions to electronic versions. **LAN(w)** should be interpreted in this way, where "printed" includes hand written and electronic versions of the same text.
[24] This is beautifully described by Carl Sagan here: https://www.youtube.com/watch?v=MVu4duLOFGY.

**A basic problem: seeing the written page** In *The Grand Illusion* (Goodman *et al* 2016), the authors comment on how our impression of seeing a whole page of text in front of us when reading is an illusion – a construction of the mind – because in fact our eyes see only a small part of the page clearly, and see nothing at all in the blind spot. Gregory and Cavanagh (2011) describe the latter:

> `` *The natural blind spot occurs where axons passing over the front of the retina converge to form the head of the optic nerve, and where retinal blood vessels enter and exit the eyeball, resulting in a hole in the photoreceptor mosaic ... Each eye has a surprisingly large blind region, about 4° of visual angle, the width across your four fingers held at arm's length. .... Surprisingly, we are normally unaware of these natural blind spots. They are either filled in perceptually (a remarkable phenomenon) or they are ignored and so not seen.*"

The predictive processing model strongly supports the first option: the brain fills in the missing text, enabled by saccades: the constant movement of the eye focus across the written pages (Dehaene 2010: 13-15, Goodman *et al* 2016) and visual sampling takes place during process (Findlay and Gilchrist 2009). This illusion of seeing a complete page when reading provides strong evidence that the predictive processing model of reading text is correct. A linear model proposing translating incoming signals from the optic nerve linearly into what we "see" simply cannot explain this process.

## 6.3 The two routes to reading

The two neural routes allowing reading as referred to previously (Section 1.2) are a direct one and an indirect one (Coltheart 2000, Rastle *et al* 2001, Taylor *et al* 2013, Danelli *et al* 2015, Buckingham and Castles 2019, Willingham 2017:57, 65). This is described by Castles *et al* (2018) as follows (page 17):

> "*The fact that word reading involves more than just alphabetic decoding is reflected in all major theories of skilled reading. .... The important point is that all of the models converge in that they represent two key cognitive processes in word reading: one that involves the translation of a word's spelling into its sound and then to meaning, and one that involves gaining access to meaning directly from the spelling, without the requirement to do so via phonology..... This dual-pathway architecture for deriving meaning from printed words is also apparent in the neural implementation of the reading system*".

In symbolic form, they are

**Dorsal (Decoding) Pathway:** {Graphemes} ➔ {Phonemes} ➔ {Morphemes},
**Ventral (Direct) Pathway:**    {Graphemes} ➔ {Morphemes}.

Only the second is readily available to people who are deaf.

Note that this is characterized by Castles *et al* as theories of *skilled* reading. Indeed they state

> *"One interesting proposal that is consistent with the characterization of reading acquisition that we have put forward is that reliance gradually shifts with increasing reading skill from the dorsal to the ventral pathway (Pugh et al. 2000; Shaywitz et al. 2002)".*

We claim rather that the direct path is also possible for young learners from the start, and indeed is a powerful 'natural' way that they begin and can continue learning to read (Gray 2013) under favourable conditions. Indeed the fact that it is possible is shown by the quote from Dehaene we give in Section 1.2, emergent literacy research evidence (Bissex 1980, GoodmanY 1992, Harste *et al* 1984, Gunn *et al* 1995, Dooley and Matthews 2009) , The Visual Word Form Area (VWFA) is used "as a word letterbox" (Dehaene 2010) but is also used for other purposes (Vogel 2012, 2014, Moore 2014, Martin 2019, Vidal *et al* 2021) so it is not uniquely associated with reading.

### 6.4 Memory Issues

Memory limitations are claimed to justify the need for an essential initial skills focus, to reach automaticity and fluency with letter-sound combinations (most recently, for the South African context, see Ardington *et al* 2020). For example, Abadzi's statement about memory are that,

> *"[in terms of] working memory capacity, we are constantly performing in a very narrow timeframe of about 12 seconds. We must recognize letters and other items within a few milliseconds, otherwise we cannot hold the messages they convey in our minds long enough to interpret them or make decisions; by the end of a sentence, we forget the beginning….Higher-order skills emerge only after the very basic skills are tied to the point of automatic and fluent performance"* (Abadzi 2006: 585).
> *"Novice readers who make conscious decisions about letters can only read small amounts of text and may have to read a message repeatedly to understand its meaning"* (Abadzi 2006:586).

The problem here arises due to focusing learner's attention on the imperative to attend to combining and memorizing the small details, which appear meaningless. Of course this will overburden working memory. Attending to meaning using various cueing systems described above, orients learners towards reading words, phrases, and sentences holistically.

These are stored in working memory as chunks, solving the problem of memory overload. Attending to combining letters into sounds should only be done when necessary in service of this process: *"…a language user engages in the process of seeking meaning through the grammatical structures. He (sic) uses the surface structure, the sequences of sounds and letters, only as signals or means of getting at, or inducing or recreating, the deep structure"* (Goodman 1982:55).

Abadzi's assumption of working memory overload (also see Adams 2001) which is claimed to restrict young learners initial focus (and which is why she claims they have to focus on the letter sounds first) is also challenged by Merlin Donald. He states that the laboratory studies that this assumption is based on look only at the lower limits of conscious experience (Donald 2001: 47). Working memory in real life is much larger than this and supports the remarkable capacity we know toddlers and young children have for grasping and memorizing new vocabulary and sayings while involved in going about their daily life.

## 6.5       The autonomous, context- free linear model

The reading model proposed *inter alia* by Abadzi (2006) and Castles *et al* (2018) is skills based, 'bottom up' and linear. Castles *et al* (2018) discuss it as follows:

> *"What does the product of successful orthographic learning look like?  First, according to Perfetti (1992), it involves having developed fully specified, rather than partially specified, internal representations. By full specification, Perfetti means that the input code is sufficient to uniquely identify the word to be read, without the necessity for discriminating between several competing partially activated candidates... in these circumstances, the correct word is specified completely by the input code, context does not need to be used to assist in the identification of the word….*
> 
> *... skilled "lexical" retrieval is effectively modular, and is only very minimally influenced by factors other than the input code".*

This says that reading does not proceed along the non-linear predictive lines that all perception uses, as we have explained above. They confirm this view by stating,

> *"Consider once again the example of the word `face'. Successful discrimination of this word from the many other words in English that differ from it by only one letter (e.g., fact, lace, fame) requires the reader to develop a very precise recognition mechanism, one that attends to all of the letters in the word and their order. Otherwise, identification accuracy and access to meaning will be compromised."*

There is no recognition here that a competent reader does indeed recognise a word by its context, even if the word is jumbled (**Section 2.4**). One can deduce the word is "face" not "lace" or "fact" or "fame" if it is in a meaningful sentence as is illustrated in **Figure 3**. One does not have to read all the letters as they claim.

This requirement of strict precision contrasts sharply with an understanding where the status of reading as a form of perception is recognised, following the same principles as all other forms of perception: missing data is filled in according to context by a predictive model (Section 2). It also contrasts strongly with what Castles *et al* themselves state later: *"Inferences need to be made beyond what is overtly stated to establish meaning within and between sentences, and need to draw on background knowledge."* Just so.

This contextual process assists in word and letter discrimination (Willingham 2017: 60-63), The non-linear hierarchical predictive model shown there is in complete contrast to this linear model. It is enabled by predictive generative processes dependent on context. This is simply not a bottom up linear reading process. Consequently Friston *et al* (2017a) state,

> *"The key thing to take from these results is that the agent can have precise beliefs about letters without ever seeing them... it is not necessary to sample all the constituent letters to identify a word. Conversely, there can be uncertainty about particular letters, even though the subject is confident about the word."*

This crucial point to note is that the core of the reading process is one which does not require  getting all the details right first. This is not needed for the communication task that is the central purpose of reading (Friston *et al* 2020).  Furthermore perception of words and letters depends on context (Rumelhart 1977).

**The Simple View of Reading** (Section 1.2) is based on a context-free linear model (see eqn. (1)). However, first, explicit decoding is only necessary for one of the two reading pathways (Section 5.3). The Ventral (Direct) Pathway functions without such an explicit process. Even though deciphering structural features is happening, it is not a letter by letter decoding process. Second, an ability to comprehend early in reading development can be constrained by decoding if reading is taught by methods orienting the learner's attention on decoding, rather than in ways based in meaning **(Section 6.3)**. That is a limitation resulting from a particular teaching method. Third, it is not clear that capacity to read jumbled words can in fact be accounted for by the SVR because of its strict reliance on decoding. But we do indeed have that ability (Section 2.4). An interactive model is far more plausible (Rumelhart 1977).

**Decoding First** The SVR is closely associated with the dominant view that decoding must take place first, as stated for example by Patael *et al* (2018): *"The ultimate goal of reading is to understand written text. To accomplish this, children must first master decoding, the ability to translate printed words into sounds."*. But they then carry on, *"Although decoding and reading comprehension are highly interdependent, some children struggle to decode but comprehend well, whereas others with good decoding skills fail to comprehend. The neural basis underlying individual differences in this discrepancy between decoding and comprehension abilities is virtually unknown."* Indeed their very careful study show that such a discrepancy is real. We suggest the resolution is that the premise is false: when reading takes place by the ventral pathway, such a discrepancy can be expected. The brain then acts in a predictive way, as discussed above.

## 6.6     Neuroscience evidence and reading: reductionist research methods

When considering the neuroscience evidence supporting either of these views, one should be very aware of the strengths and limitations of the evidence provided. Because evidence for skills-based reading models is based on a reductionist view of brain function, it necessarily incorporates the limitations of that view. More specifically, books like Deheane (2010) have major limitations in terms or providing evidence regarding the reading process. They study parts of what is involved in reading, but not the integral process of meaningful reading. Thus they can only provide evidence about isolated aspects reading, not how they are integrated to enable the process as a whole.

Even then the studies are really limited: Castles *et al* (2018) state *"most of the work on spelling-sound relationships has been conducted with monosyllables; researchers are only just beginning to consider spelling-sound relations in letter strings with more than one syllable"*. This is hardly sufficient to determine how meaningful language works. Related to this, there is a lot of data on reading nonsense words and phonemes. This gives no data on the integral process of reading meaningful text. That aspect is missed by all brain imaging studies which look only at how phonemes or pseudo-words are processed.

An example of such limitations is a study by Cattinelli *et al* (2013), who performed a new meta-analysis based on an optimized hierarchical clustering algorithm which automatically groups activation peaks into clusters. They focussed exclusively on experiments based on single words or pseudowords from the following four classes of tasks: reading, lexical decision, phonological decision and semantic tasks. But you can't do a real semantic task based on single words or pseudo words. This kind of study can only be useful to determine isolated parts of the reading process. It should not be taken to give information on the actual reading process. It simply does not have the necessary data and should not be treated as if it does.

**6.7      Neuroscience evidence and reading: holistic research methods**

Extensive work has been done to put the study of real reading on a scientific basis, as summarised in Flurkey and Xu (2003) and Flurkey *et al* (2008). The latter state,

> ``*The emerging concepts from [current] research clearly indicate that the higher cortical structures control the transmission of information from the deeper structures. This interpretation is contrary to the classical teaching, in which deeper sensory relay stations determine what will eventually reach the cortex. The emerging view has profound implications for psychological models of mental life. Whereas the classical neuroanatomic view is most consistent with a bottom-up, information processing model, the emerging view supports an interactive, constructivist model. The cortex either promotes or inhibits the very input being transmitted to it from the eyes, ears, and other sensory receptors. ... the cortex selects evidence to confirm or disconfirm its predictions. It anticipates what will be seen and heard using knowledge stored in memory. Both this new neuroanatomical view and its psychological reflection are consistent with a transactional sociopsycholinguistic model of reading.* "

This is precisely the predictive processing view discussed above. It is supported by evidence as follows:

First, *eye tracking studies* Evidence comes from eye movement analysis of fixations, omissions, and backtracking. Since the most conspicuous motor behavior in silent reading is eye movement, studying it allows us to "see" the silent reading process (Flurkey *et al* 2008, Seidenberg 2017:62-70). We do not in fact read every word (Goodman *et al* 2016). Not all words are read: some are skipped. Visual sampling takes place during text reading (Findlay and Gilchrist 2003).

Second, *miscue analysis*. When combined with miscue analysis from oral reading, it is clear that cortical instructions tell the eyes where to look for cues from the signal, lexico-grammatical, and semantic levels of language - the three cueing systems (Flurkey *et al* 2008, Goodman *et al* 2016).

Third, g*arbled words and phrases* The way that we can read sentences when words are mis-spelled or missing, or when letters are re-arranged within a word (Section 2.4) or grammar is wrong is strong evidence of how reading works in a contextual way.

Fourth, *letters are sometimes identified in a top-down way,* based on the what the probable word is (Willingham 2017: 60- 63; and see **Figure 3**).

Fifth, *inferring meaning and pronunciation*. We often have to infer in a top-down way what part of speech a word is and what it means through context (e.g. "plane", "flies"). Sometimes the way a word sounds may depend on context (e.g. "wound" has multiple meanings and pronunciations). This is a common feature of many languages, irrespective of orthographical features.

Sixth, *brain imaging studies.* Flurkey *et al* (2008) comment that the subjects in the various brain imaging studies of reading at the time they wrote had not been given phonological processing tasks embedded in a context that requires meaning construction, nor have they even considered imaging studies illuminating the effect of home reading programs on neural development. Such studies have recently been initiated by J S Hutton and co-workers, who have applied MRI studies to better understand the influence on structural and functional brain networks of young children in home reading environments supporting emergent literacy. They are obtaining information on neural processes related to actual reading processes,[25] and the accompanying skills and attitudes which develop; for example fluent reading was found to be supported by executive function areas. (See Horowitz-Kraus and Hutton 2015, Horowitz-Kraus *et al* 2017, Hutton *et al*. 2015, 2017, 2020).

All this emerging data provides strong evidence for the meaning-construction view of reading. The transactional socio-psycholinguistic character of reading is an instantiation of the non- linear, integrative memory-prediction model of brain function discussed above (Section 2). Following on Sherman and Guillery (2006), Flurky *et al* (2008) emphasize the role in these processes of thalamo- cortical circuitry, in agreement with Alitto and Usrey (2003).

## 7    Oral and written language learning

What about the nature of learning to understand and use oral and written language? The similarities between the processes involved in oral speech and written communication suggest that there should be important similarities in the conditions babies require to learn to listen and speak, and young children require as they learn to read and write (Holdaway 1979, Cambourne 1995). Without role models who interact with them and surround them with demonstrations of language being used for various purposes, babies would not have the social context that supports and shapes oral language development (Hof 2006). The same applies to learning to read and write. It is this which leads to understanding and supporting the growth of literate environments and reading culture development in providing the conducive conditions for literacy learning.

### 7.1   Basic principles of language learning

The following can be claimed to be basic principles underlying learning both spoken and written language.

*a)*     **Constructing and Conveying Meaning** In learning to speak, the foremost thing babies have to learn is that *spoken words convey meaning and emotion and information and stories*[26] (this is **LAN(o)**, **Section 5.1).** This empowers the drive to understand and to listen and attempt to speak, as they try to make sense of and predict the world around – as well as  the need to communicate with significant others. Similarly, in learning to read and write, the foremost aspect toddlers/children have to learn is that *written words convey meaning and emotion and information and stories* (this is **LAN(w)**, **Section 5.2).** This too powers the intrinsic motivation to explore and communicate in ways which include using print. It is closely tied in to the key process of learning to read minds, which as stated in a very useful paper by Heyes and Frith (2014), is like learning to read print.

---

[25] Friederici (2017), in particular pp.121-141, presents such studies in the case of oral language.
[26] We mean here stories in their broadest sense, incorporating the narrative form.

b)      **Joint social processes** Learning to speak and understand and learning to read and write are both joint socially based processes involving attempted efforts and feedback, and with a strong affective component. This means each is as 'natural' as the other (Goodman and Goodman 2013, and Section 4): neither *has to* take place in a formal educational context (Bissex 1980, Taylor 1983, Bloch 1997). The processes are culturally shaped, with the carer/teacher expectations themselves being shaped by the adult's own prior experiences and understandings (Heath 1983).

c)      **Successive approximations** Both these socially based processes of learning involve successive approximations enabled by the specifics of socio-cultural and educational contexts the child encounters. She learns phonological and phonetic principles: the relationships between sounds and meaning in the case of spoken language, and graphemic and alphabetic principles of written language when writing is based in letters drawn from an alphabet. In each case learning is a process of observation, experimentation, and successive approximation to reach the correct form (Heyes and Frith 2014), with errors corrected by feedback through repeated demonstrations of conventional speaking and writing.

d)      **Building on existing strengths** When they learn to speak, read, and write, children draw on all of their learning strengths to move from the known to the unknown (Bruce 2015). This includes their understandings, knowledge, and uses of oral language, its vocabulary, metaphors, and grammar in one or more languages, as they begin to include written language in their communicative repertoire (Au 1980). Thus a major predictor of success in learning to read is the presence of an already reasonably well developed spoken language and vocabulary in the same language.

e)      **Motivated to engage** High motivation to learn and practice is a central aspect of both oral and written language learning. Making meaning of the great complexities of written language needs high and consistent levels of motivation and engagement with texts, affects comprehension (Wigfield *et al* 2016). A child's self - confidence, beliefs, values and goals, as well as sense of autonomy and interest all play a significant part (Barber and Klauda 2020), if intrinsic motivation continues to be encouraged beyond the early years, activities related to positive achievement are greater than with extrinsic rewards (Ryan and Dechi 2009).

**7.2    Learning oral language**

How does learning oral language take place?  Shaywitz (2003) claims that oral language does not have to be taught because learning to speak is a natural process. This claim is widely accepted now by policy makers, academics, and language specialists as being based in undisputable scientific evidence. But why is it natural, given the complexity of the task? We suggest that this is because it takes place through the predictive processing kind of interaction emphasized in this article, which is one of trial and error followed by feedback and correction. It involves an informal and superbly effective teaching process because babies have the kind of conditions they require to learn when family members speak constantly to and around them. Babies want to understand and be able to express themselves too; caregivers and others have high expectations that babies are capable of learning to listen and speak, and talk to them as if they already understand as they try to meet their needs and moods. Castles *et al* (2018) state :

> **LEARN(o):** *"If a child is exposed to a rich oral-language environment, that child will almost certainly learn to understand and produce spoken language."*

Such an environment involves enormous numbers of everyday verbal interactions, initially with carers, who guide the ongoing reciprocal interaction, experimentation, practice, and play as babbling emerges. Over time, and with ever better approximations of the accepted speech of the particular community, it becomes the appropriate form of conventional spoken language. This has three dimensions: Firstly the child must learn the motor control involved in speaking: shaping the tongue and lips, controlling breathing, and so on. Secondly she must learn to apply phonological principles which transform sounds into words and sentences. Thirdly she must learn how and when to use the grammatical, lexical, and cultural and linguistic conventions to convey the meanings of her speech community.

As we have intimated, from our viewpoint, the key issue overlooked by many is that *this IS a teaching environment.* It is an *informal* teaching environment (Lave and Wenger 1991, Rogoff *et al* 2016), involving the necessary conditions which support learning (Cambourne 1995). In terms of the discussion in the next subsection, this is an apt example of "natural learning" (Holdaway 1979) corresponding to the need to create meaningful, holistically oriented teaching environments.

### 7.3   Learning written language

How does learning to use written language happen? It can take place in both informal and formal teaching environments. It can be oriented to be either a skills based process, emphasizing the parts first and then building them up to create wholes, as summarised in **Figure 1** or a meaning-based process, emphasizing engaging with and composing whole texts while also appreciating and attending to the contributing parts, as summarised in **Figure 2**. Reductionist skills-based approaches insist on getting the details right first before moving on to use reading and writing for authentic reasons (hence the widely used phrase, 'learn to read, then read to learn').  Holistic, meaning-centred approaches support learning through successive approximations towards conventional reading and writing.

According to Castles et al (2018) *"The fundamental insight that graphemes represent  phonemes in alphabetic writing systems does not  typically come  naturally to children. It is something that most children must be taught explicitly, and doing so* is important *for making further progress in reading."*

The key issue here is the phrase "come  naturally to  children". What is understood as natural depends crucially on cultural context (Rogoff 1990). If you live in a highly literate environment that uses and displays as normal writing in a language you are comfortable using, what comes naturally is quite different than if you do not. And what does "taught explicitly" mean? If a mother teaches her child to spell her own name on a sheet of paper, is that explicit teaching? We would suggest yes. It is not part of an explicit teaching program: but it is teaching nonetheless, just as is being taught to say her name in the case of spoken language.

 It is just as natural in both cases, given appropriate conditions. In other words, to learn to read children have to read and be read to (Smith 2004), while to learn to write, they have to write – and read too - as potential authors, guided by teachers and others who write themselves so that as they begin to write, they come to see themselves as writers (Smith 1983).  Infants and young children struggle to begin a 'natural' process of learning if they are in settings with few relevant role models using written language in ways which interest and  draw them in as newcomers to a cultural practice (Rogoff *et al* 2016).  They are enabled to begin this process effectively by observing and joining in voluntarily to personally relevant activities involving writing and print in relevant languages, be these in homes, community settings, or school contexts.

This kind of informal learning is illustrated by a Polish colleague who tells of his induction into reading as follows: he had a brother who was 4 years older than him, and at that time, school started when children were 7 years old. He was 3 when his brother started to learn to read, sitting in their common room at a small table in the middle of the room. The older brother would be reading the letters and words aloud, running his finger below the line of print. Our colleague would be kneeling on a chair at the other side of the table following his brother's finger. Within a year (by age 4) he had learnt to read fluently - upside down! Only later did he learn to read with the 'normal' orientation. No formal skills teaching occurred in this self-motivated, socially contextualised process. This is one of many cases that demonstrate the successful nature of informal teaching; it is not essential to have formal teaching in order to learn to read.

While we are not in any way claiming here that teaching reading is not necessary, it is well documented that children can learn the fundamentals themselves under appropriate conditions (Clark 1976, Buckingham and Castles 2019). Indeed up to 5% of children are "precocious readers" who do this (Olson *et al* 2006)**.**

In parallel to **LEARN(o)** in Section **6.2,** the following is plausible:

> **LEARN(w):** *If a child is exposed to a rich, contextually relevant written-language environment, which involves that child in regular, satisfying reading and writing interactions with significant others, including shared attention to the details of the process, and constant positive feedback, that child is highly likely to learn to understand and produce written language.*

### 7.4  The similarities between learning spoken and written language

The predictive processing viewpoint, and more generally the way perception functions as discussed in Section 2, can be claimed to support learning both processes in neural terms, based in the statistical pattern recognition properties of neural networks. discussed in Section 5.  Consequently, Our view is that learning spoken and written language are underpinned by very similar processes, as indicated in the box below (Bloch, in Ellis 2016:448):

> **Language is listening, speaking as well as reading (including braille), writing and signing**
>
> | Baby learns to speak | Baby learns to read-write |
> |---|---|
> | • Hears, sees/ experiences people who speak (role models) | • Hears, sees/ experiences people who read – write (role models) |
> | • Expresses and communicates as she learns | • Expresses and communicates as he learns |
> | • Learns why she listens and talks at same time as she learns how | • Learns why he reads - writes at same time as he learns how |
> | • Has shared interactions | • Has shared interactions |
> | • Is included, heard, encouraged, praised - connects emotionally | • Is included, heard, encouraged, praised - connects emotionally |
> | • makes 'mistakes' - speaks immaturely(babbles) and plays with sounds. | • makes 'mistakes'- reads/writes immaturely (pretends to read, does emergent writing). |

### 7.5 Implications of an integrative view for progress in meaningful reading

Considering the integrative body of neuroscience discussed above, what could detract from and what could support children learning to read and write with meaning? Whilst in this paper we don't detail early literacy teaching methods, and acknowledge the huge body of existing expertise in this regard, we make the following general points:

A major issue for learning effectively exists in multilingual print scarce settings, like South Africa, with de facto language policies that move to teach from African to ex-colonial languages after only three years schooling (Mkhize and Balfour 2017, Bua Lit 2018). Here the potential for compromised understanding already exists to such an extent that it can feel normal. This makes it easy to accept that teaching and assessing reading doesn't involve comprehension until later. The drive to search for meaning can thus be minimized, deflected, or hidden when the broad initial orientation is towards separate skills teaching, with phonics automaticity and fluency must be mastered as an initial imperative.

This is particularly so if access to compelling fiction and non-fiction material in preferred languages, is absent or positioned as supplementary, and there are few or no reading and writing role models to interact with. Limited vocabulary books, which have been 'levelled' are used far more[27] than materials which stimulate curiosity, challenge imaginations, and encourage inference and problem solving. Such materials don't necessarily hinder the progress of children who engage elsewhere with emotionally

---

[27] A wonderful diatribe against such books is given in the section on Education in *Let us Now Praise Famous Men* by James Agee (Houghton Mifflin 1988): pages 289-307

satisfying texts which build vocabulary and language knowledge as they conjure awe and excitement. But children who have to rely on school for such motivation and enrichment may wait for so long that they give up and never get what they need.

Though learning letter sound combinations and relationships is integral to learning to read and to write in alphabetic languages, we contest the validity of teaching it in prescriptive ways, dissociated from the wider fields of meaning and personal relevance and agency. Phonics based methods are acknowledged to possibly delay the relation to meaning until automaticity and fluency have been attained (Seidenberg 2020). The interim learning is often low level and mind-numbing; it is highly questionable whether this can contribute to the much desired recipe for success. Telling children that this will change once they have learned to read does not necessarily help: the experience of meaninglessness is real.

Apart from not fully assessing elements which indicate reading progress, assessments using non-words (Castles *et al* 2018:19, Bua Lit 201) and meaningless phonemes, such as the widely used Early Grade Reading Assessments, reinforce the message that reading is not related to anything personally useful or interesting. Again this can be highly demotivating.

In contrast, an orientation which provides a relevant base for meaningful learning emphasises the value of children's languages, emotional and personal knowledge and connections from home and community.. From this place of respect and belonging, stories which can be fictional, factual, or historical stimulate imaginative engagement. Teachers can learn to teach phonics and other skills as and when needed by children as they read and write (**Figure 2**). Regular, interactive experiences with worthwhile[28] texts, involving plenty of teacher read alouds and conversations with children to motivate and stimulate imaginative thinking and use of language, should begin early and continue to be supported and overtly valued.

Horowitz-Kraus and Hutton (2015) confirm this by stating[29]

> *"Children utilising imagery during stories listening will have greater success in reading later in life, which is consistent with findings suggesting that better utilisation of imagery during stories listening improves comprehension. Studies citing quotes of children's experience when listening to stories confirm that imagery supports this process , even more intensely for stories without pictures , perhaps via more intense activation of the visual association cortex"*

Castles *et al* (2018) states,

> *"The single most effective pathway to fluent word reading is print experience: Children need to see as many words as possible, as frequently as possible … statistics point to the huge value of fostering a love of reading in children and a motivation to read independently.*"

We agree and suggest that an assumption in this statement needs to be overt: a love of reading is made possible when teachers orient themselves to appreciate the importance, legitimacy, and power of becoming well-informed, interactive role models who read aloud well and frequently to children, encouraging curiosity, imaginative and critical thinking, and real conversations about what's being read. This ought to be normalised as the essential orientation for all early literacy teachers. Even in the highly print saturated settings of the UK, reading for pleasure has declined, (National Literacy Trust 2019) and fresh evidence is emerging as to the rich literacy teaching benefits of ensuring that teachers themselves read for pleasure and indeed are readers in their own right (Cremin et al 2008, Cremin 2020). This deceptively simple notion helps teachers to awaken the desire to read in children by harnessing the pleasure, enriched language, and other opportunities literature holds for learning (Krashen 1989, Arizpe and Styles 2016, McQuillan 2019, Bloch 2015). It is also the springboard from which to support teachers to encourage children to apply strategies which include multiple cueing systems, a focus on authentic composing and writing and to consider related assessments which address multiple dimensions of literacy.

---

[28] We use the term 'worthwhile' to reiterate the benefits of teachers and teacher educators engaging in an ongoing investigation of books, with discussion about what 'worthwhile' means in diverse cultural contexts. It points to the extraordinarily important role adults have in curating the texts children encounter, and also to their observing and consequently learning from and about the children who explore the books.

[29] In stark contrast to Dehaene (2010)

### 7.6 Conclusion

Far too many young children's literacy learning opportunities are being compromised daily by the increasingly wide acceptance of a restricted, reductionist body of neuroscience evidence as being the true and unquestionable basis for teaching reading. The following statement referring to teaching in South African schools, summarises how this view is interpreted for teachers:

> "*Unlike learning to speak, decoding does not come naturally; it is a method that must be taught systematically. It is important to emphasize that reading is produced by the product of vocabulary and decoding: If one has a perfect vocabulary but has not been taught the method of decoding one will not be able to read at all. Letter recognition and phonemic awareness are mastered through systematic teaching and consistent practice. This leads to the next stage of reading acquisition: word recognition. Through practice and appropriate progression from simpler sounds and words to more complex ones, word recognition becomes established leading to the next phase of reading acquisition: fluency. It is only once decoding and word recognition have become fluent, even to the point where it becomes automatic and unconscious, that it is possible to reach the ultimate goal of reading comprehension*" (Taylor et al 2019: 20).

What allows children to achieve this perceived initial mastery? They continue (Taylor *et al* 2019: 21).:

> *"In order to learn the basics of decoding, a child requires a teacher who is present, capable and motivated to deliver systematic reading instruction. In order for decoding to become fluent a child requires suitable graded materials and the discipline (perhaps imposed) to practice a lot."*

This rigid and foreboding vision of what it could mean for teachers in over-crowded and under resourced classrooms to (perhaps impose) discipline on young children to practice their graded materials (if they even have these) is a depressingly common consequence of relying on this reductive model.

We have contested this vision with the body of integrative neuroscience which supports the view that all understanding is contextual. Learning starts at birth: young children's brains are capable of handling complexity and learning meaningfully from the outset, outside of exceptional cases. This is confirmed by the body of early literacy evidence detailing young children's emergent reading and writing prior to formal schooling (Whitmore *et al* 2004, Nutbrown 2018, Carroll *et al* 2019, Teale *et al* 2020 ). Observations of young children reveal much time and effort spent with voluntary skills practice when these skills interest children and form part of play or other authentic purposes - and this includes children from poor communities (Sibanda and Kajee 2019, Bloch and Mbolekwa 2021).

School literacy teaching should continue to develop such foundations and build on them, in ways which respond sensitively to children's ongoing meaning-making endeavours. Integrative neuroscience offers evidence to support this, implying the value of teacher education programmes which problematise narrow interpretations of the science (Hoffman *et al* 2020),  renewing attention to and research on teaching approaches and  methods currently eschewed or  straight jacketed to fit reductive neuroscience understandings. All teachers, especially those in under-served settings, need overt, systemic support to provide children in their first years of formal school with the kind of culturally responsive, rich learning opportunities that are currently afforded in reasonable quality only to children from affluent communities. Among many others, Cambourne (2000, 2017) and Whitmore and Meyer (2020) provide solid foundations for this endeavour.

**Acknowledgements :** We thank Eva Bonda, Thomas Parr, and Tina Bruce for helpful comments, Mark Solms for providing Figure 3, Mandy Darling for (re)drawing the figures, and Roland Eastman for extremely helpful discussions that shaped Figure 1, as well as regarding details of the text. We thank two referees for comments that have materially improved the paper.

Cape Town 2021/03/25

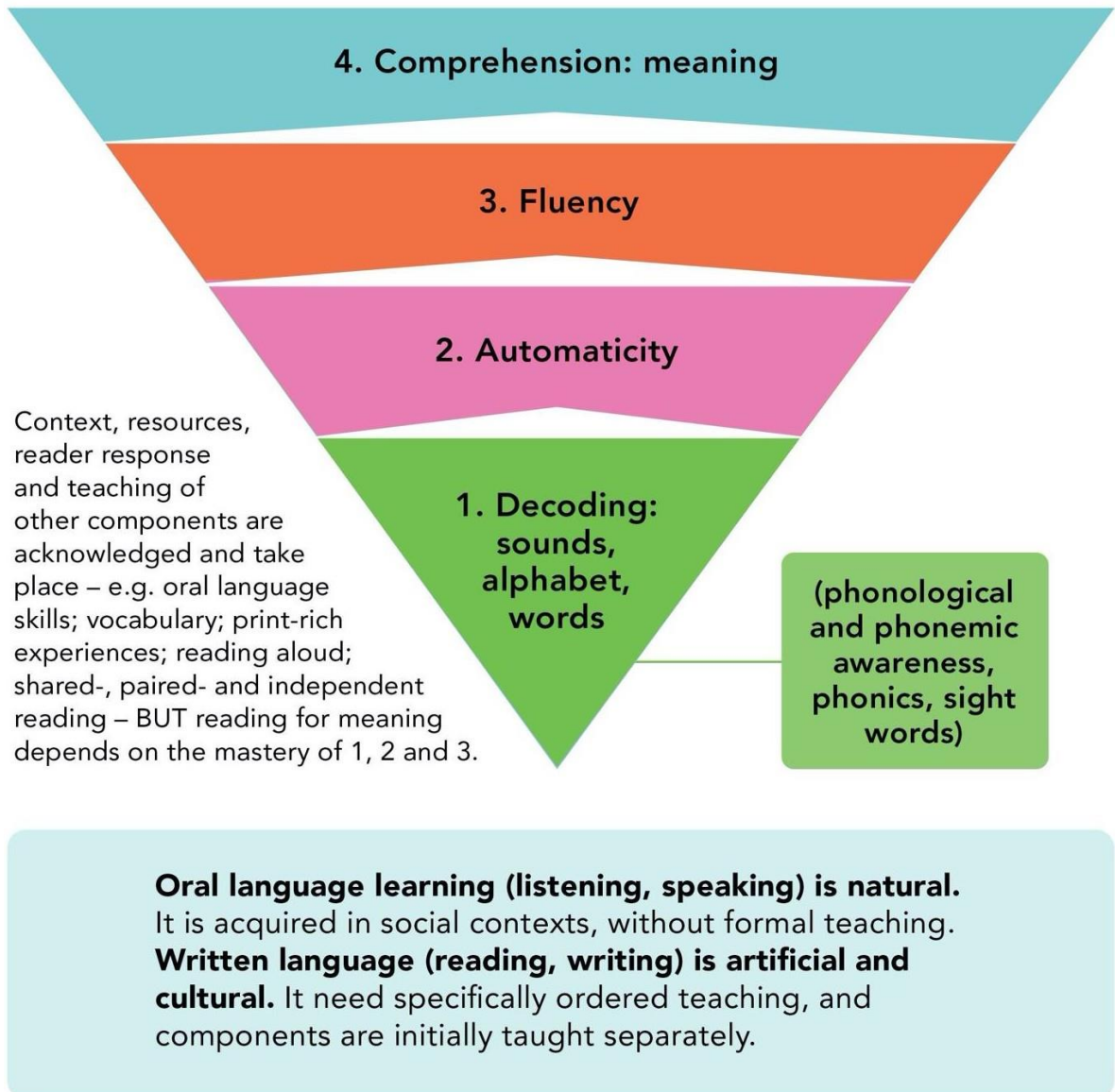

**Figure 1 Skills based model of learning to read.**

Figure source: Carole Bloch.

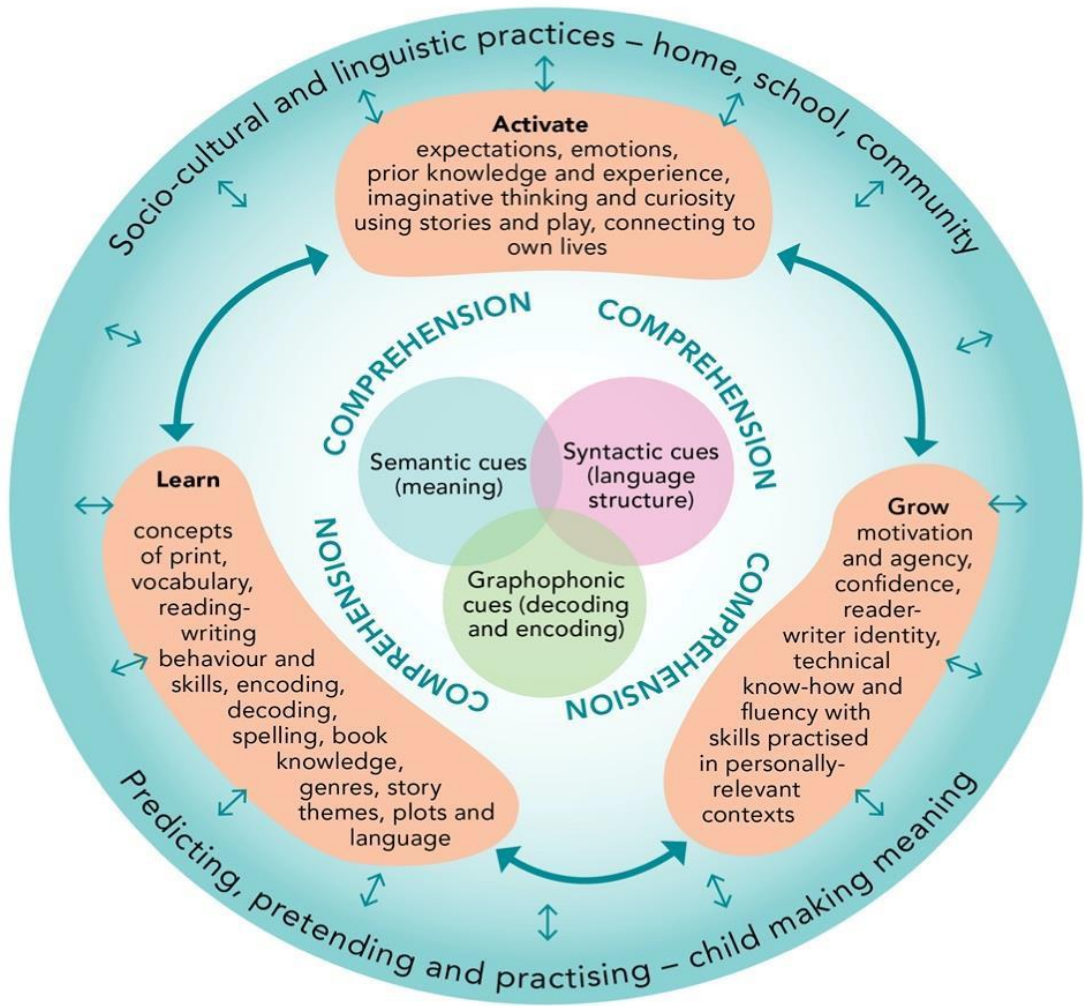

**Figure 2 Meaning-based model of learning to jointly read and write.**

Figure source: Carole Bloch

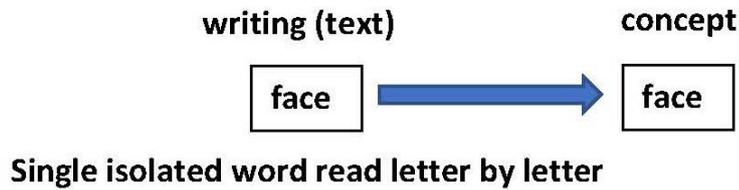

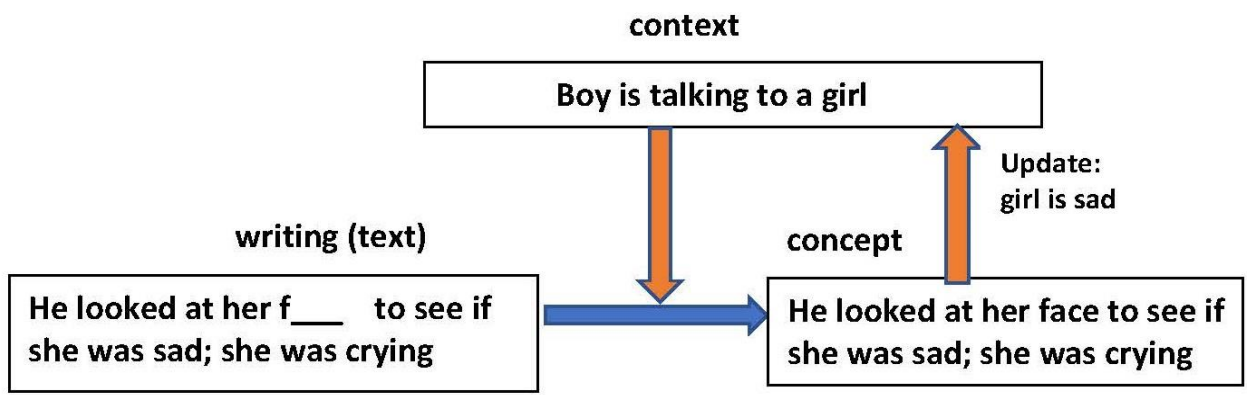

**Figure 3 The linear reading process envisaged by the Simple View of Reading (A), and how contextual information enables word identification in a contextual way (B).** *When the context is updated in response to the information gained, one has a model of the predictive processing understanding of the reading process. The closed loop makes it non-linear. The letter ambiguity Castles et al (2018) discuss (`fact' or `face'?) can be resolved in this way without reading every letter. The outcome is an interactive model of reading (Rumelhart 1997) in agreement with Seidenberg et al (2020).*

Figure Source: George Ellis

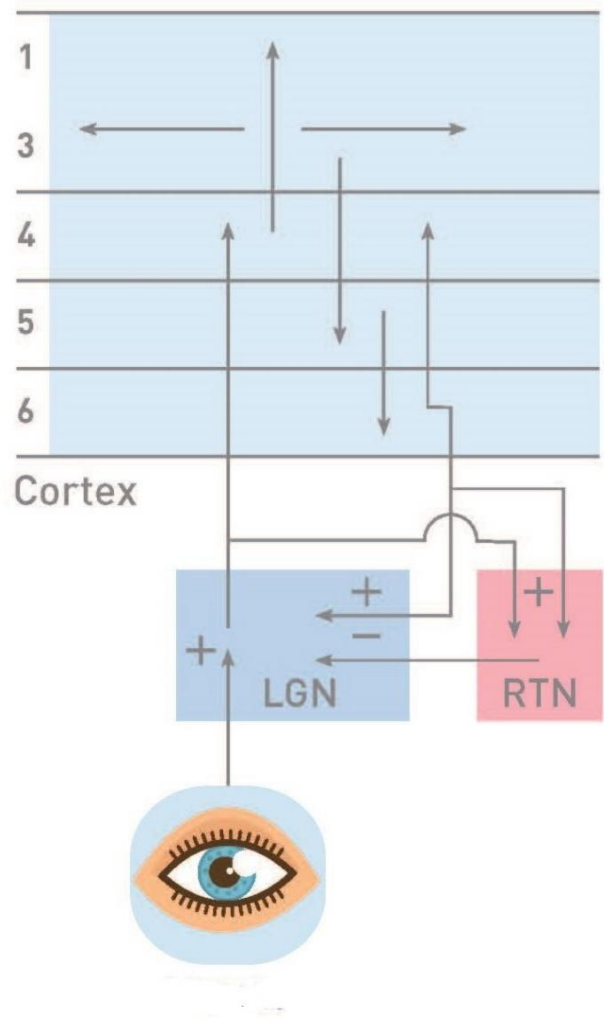

**Figure 4 Corticothalamic circuitry for the visual system.** *Information flows from the eyes via the optic tract to the Lateral Geniculate Nucleus (LGN) in the thalamus and then via excitatory projections to level L4 in the visual cortex and on to levels L3-L1. Predictive information flows down from L3 to L5 and L6. Neurons in L6 send excitatory feedback to the thalamus and the reticular nucleus (RTN). The feedback axons terminate on relay neurons in thalamic relay nuclei, as do inhibitory projections from the RTN*

Figure adapted from Alitto and Usrey (2003).

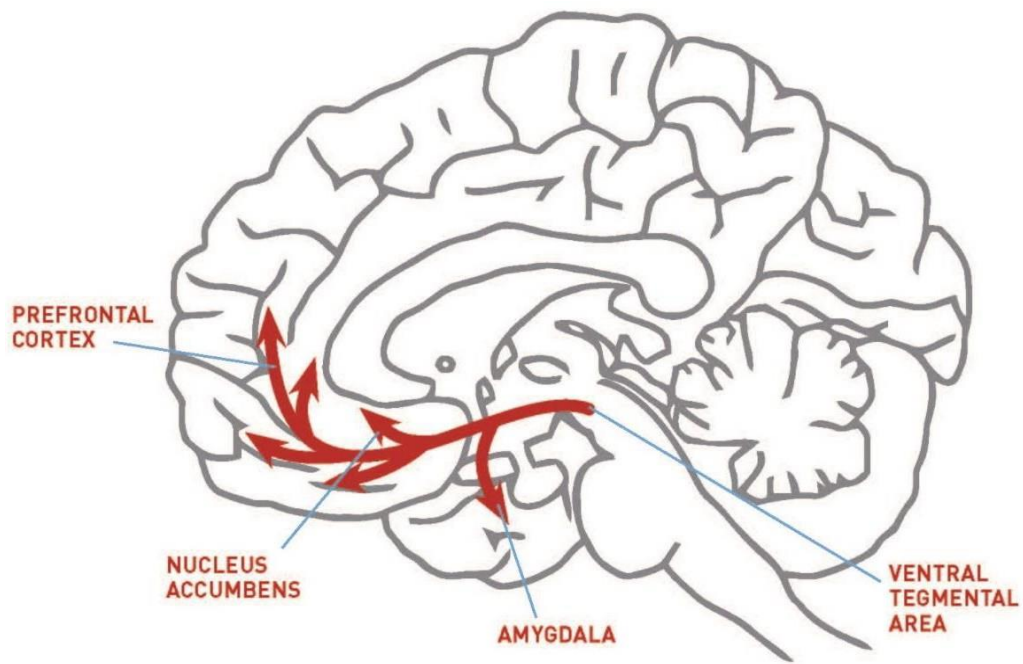

**Figure 5** *The SEEKING system is one of the ascending systems that project diffusely to the cortex from nuclei in the excitatory systems, conveying neuromodulators such as dopamine and epinephrine to the neocortex. These reticular activating systems underlie Gerald Edelman's Neural Darwinism* (Edelman 1987) *as well as Panksepp's primary affective systems* (Panksepp 1998.)

Figure source: Mark Solms.

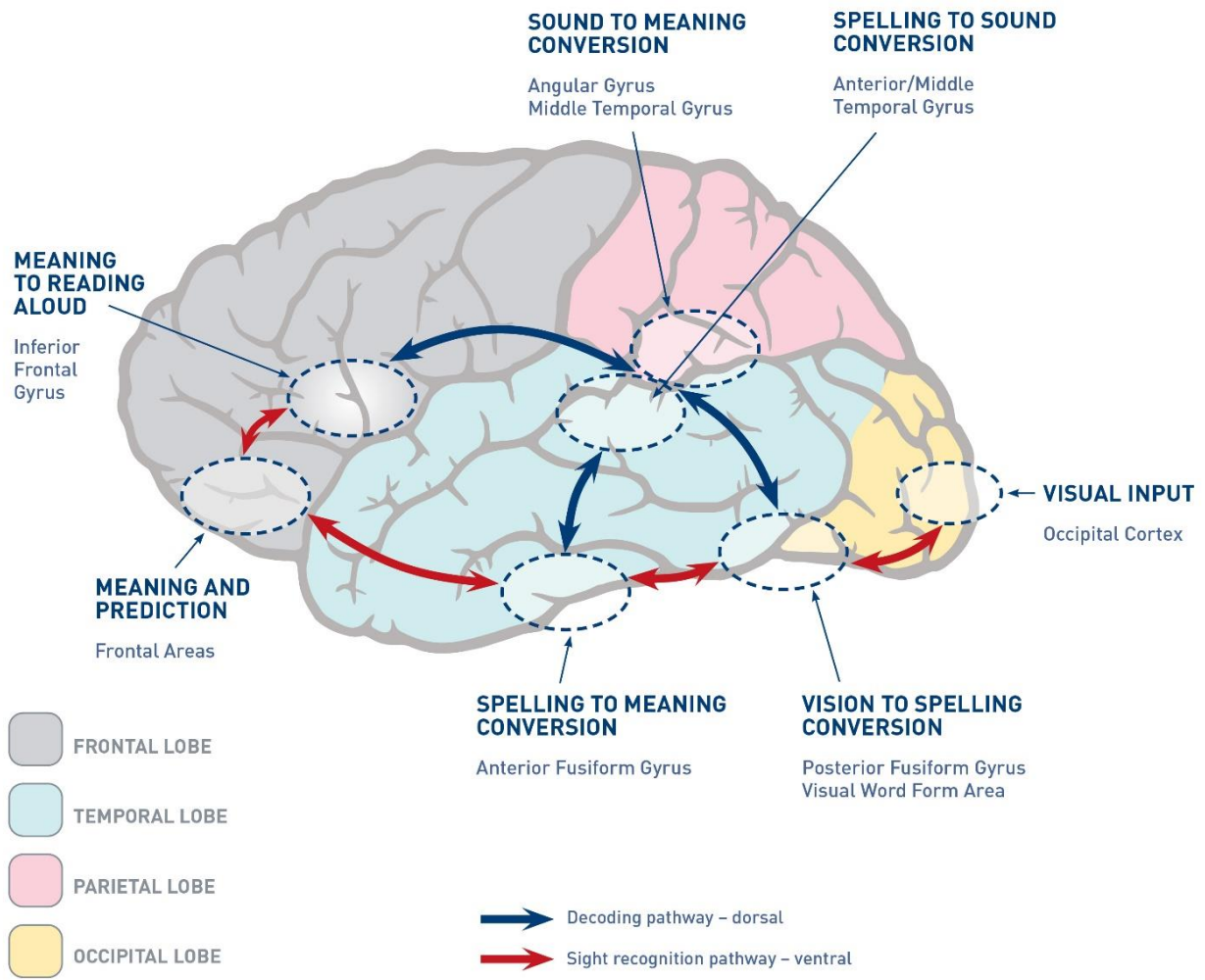

**Figure 6 Brain pathways associated with reading**. *A dorsal pathway underpins phonologically mediated reading, and a ventral pathway underpins direct access to meaning from print. Many further cortical areas will be involved when meaningful reading occurs, for example reading stories with an emotional impact, and the brain engages with that meaning in its social context. Standard neuroimaging studies do not emphasize these further areas because they do not deal with the reading of meaningful texts For analogous diagrams in the case of oral language, see* Friederici (2017:pp 107,109,124,128,135).

Figure adapted from Taylor *et al* (2013), Rastle *et al* (2001), and Kearns *et al* (2019), under the expert guidance of Professor Roland Eastman (former Head of the Neurology Department, University of Cape Town).